\documentclass[11pt,a4paper]{article}

\usepackage[margin=2.5cm]{geometry}
\usepackage{amsmath,amssymb}
\usepackage{booktabs}
\usepackage{graphicx}
\usepackage{caption}
\usepackage[colorlinks=true,linkcolor=blue,citecolor=blue,urlcolor=blue]{hyperref}
\usepackage{xcolor}
\usepackage{natbib}
\usepackage{setspace}
\usepackage{authblk}
\usepackage{placeins}
\usepackage{pdflscape}
\usepackage{multirow}

\newcommand{\BIM}{\mathrm{BIM}}

\begin{document}

\title{Auditing the Global Carbon Budget:\\
Exploring the 2024--2025 Vintage Shift}

\author[1]{Mikkel Bennedsen}
\author[1]{Eric Hillebrand}
\author[2]{Siem Jan Koopman}
\affil[1]{Department of Economics and Business Economics, Aarhus University,
  Aarhus, Denmark}
\affil[2]{Department of Econometrics, Vrije Universiteit Amsterdam,
  Amsterdam, The Netherlands}

\date{\today}

\maketitle

\begin{abstract}
The Global Carbon Budget (GCB), the community reference dataset for
the carbon cycle, is reissued annually.  The 2025 release introduces
several adjustments to the published series that we compare with prior releases starting in 2017.
On a common $1959$--$2016$ sample, the mean of the GCB budget imbalance
jumps from within $\pm 0.17$~GtC/yr of zero for every vintage
$2017$--$2024$ to $+0.61$~GtC/yr in 2025, the only vintage whose
$95\%$ confidence interval for the imbalance mean excludes zero.  The
size of the imbalance changes much less: its mean absolute value rises
from $0.61$ to $0.76$~GtC/yr.  It is the mean, the quantity the budget
identity constrains, that moves.  We document and explore this shift in
two ways. First, we conduct a model-free analysis, where we
attribute the shift to a new adjustment that places
the published land sink $0.40$~GtC/yr below its ensemble mean (the
average of the underlying models), a smaller adjustment in the ocean sink in
the opposite direction, and a change in the composition of the bookkeeping ensemble.  Second, we consider the dynamic statistical GCB model of \cite{BHK2023}, augmented
with climate covariates. Its parameters are estimated for every GCB vintage 2017--2025.
The coefficients of atmospheric concentrations in the sink equations
shift in opposite directions on the 2025 issue, mirroring the model-free
findings.  A constant in the budget equation, statistically unnecessary
in every vintage from 2017 to 2024, is required in 2025 and is estimated
at $-0.59$ $(0.09)$~GtC/yr. There is a persistent drifting imbalance across the
entire sample in the budget equation.  Each of the three adjustments is documented in the 2025 release and rests on evidence
about the component it corrects. Their joint effect is a budget that
closes over the last ten years and carries a mean imbalance of
$+0.61$~GtC/yr over the full record. We argue that this cost to the full
sample outweighs the gain on the last ten years.
\\[6pt]
\end{abstract}

\newpage
\tableofcontents
\newpage

\section{Introduction}
\label{sec:introduction}

The Global Carbon Budget \citep[GCB;][]{GCB2025} is the
community reference dataset for analyses of the carbon cycle.  It is reissued each year and
brings together (1) fossil fuel emissions (and cement carbonation sink), (2) land-use change
(LUC) emissions from three to four bookkeeping models, (3) terrestrial sink estimates from around twenty dynamic global vegetation models (DGVMs), (4) ocean sink estimates from an ensemble of global ocean biogeochemistry models (GOBMs) and
observational fCO$_2$ products, and (5) changes in atmospheric concentrations of CO$_2$.  
The five components are required to satisfy the carbon budget identity, and each annual release documents the adjustments newly applied to the
component series.

The 2025 release lists the changes in \citet[][Table~3]{GCB2025}:
a ``Replaced Sinks and Sources'' (RSS) bias correction to the DGVM land
sink, an underestimation adjustment to the GOBM ensemble mean for the
ocean sink together with a correction for fCO$_2$ products, and a
transition of the bookkeeping ensemble to three models (BLUE, OSCAR,
LUCE). The bookkeeping land-use estimate of \citet{HoughtonCastanho2023} was dropped from the
ensemble.  This paper documents the cumulative effect of these
revisions on the published budget relative to earlier vintages.

The raw GCB budget imbalance (BIM) is the residual that remains when the
five published components are combined according to the carbon-budget identity.   
Comparing BIM across the GCB releases from 2017 to 2025 shows that, on the
common 1959--2016 sample, the mean of this residual remains
within $\pm 0.17$~GtC/yr of zero for every GCB vintage from 2017 to
2024. It jumps to $+0.61$~GtC/yr in the 2025 vintage. The 2025
vintage is the only one whose $95\%$ confidence interval excludes
zero.  \citet{GCB2025} reports the $1959$--$2024$ imbalance as
$0.5$~GtC/yr on average and describes it as small, which the paper reads
as evidence of a coherent community understanding of the budget
(Sect.~3.9.2).  We obtain the same number.  The paper does not set it
against earlier vintages, where the same statistic is within
$\pm 0.17$~GtC/yr on a common sample.

The 2025 data compilation follows a reanalysis of the
carbon budget published separately by \citet{FriedlingsteinNature2026} and
discussed in two accompanying commentaries
\citep{McKinley2026,Jiang2026}.  The reanalysis states its motivation as a property of
the GCB~2024 budget imbalance: over the full 1959--2023 record the
imbalance carries a small but statistically significant negative trend
($-0.14 \pm 0.04$~GtC/yr per decade, $P=0.003$), which the authors
argue impedes the interpretation of trends in the individual sink
components.  The reanalysis assembles a set of component
corrections, each within one standard deviation of the original
component and so individually insignificant, and evaluates them against
the imbalance.  It reports that they bring the imbalance over the recent
decade (2014--2023) to near zero and render the 65-year trend
insignificant, a result the commentaries describe as a rebalanced budget
that closes the carbon mismatch.

The 2025 release revises the same three components.  Its stated basis is
not the imbalance but published evidence of bias in each component:
constant biomass densities in the bookkeeping models, a land sink
computed at 1700 forest cover rather than at observed forest cover, and
an ocean sink computed at the temperature at measurement depth rather
than at the temperature of the surface layer where the gas exchange
takes place.  We take up that evidence in
Section~\ref{sec:discussion}.

Table~\ref{tab:bim_three_vintages} reports the imbalance mean and trend
under the three data sets. On the 2024 vintage, the imbalance mean is indistinguishable
from zero over the full 1959--2023 sample ($+0.00$, $P=0.99$) and is
non-zero only in the last decade ($-0.44$), alongside a trend that is
statistically significant but numerically small.  The reanalysis in \citet{FriedlingsteinNature2026} lowers the recent-decade mean toward zero but in doing so lifts the full-sample mean to $+0.18$, while removing the trend.  The
GCB 2025 vintage in \citet{GCB2025} goes further, driving the
last-ten-year mean to $+0.02$, but at the cost of a full-sample mean of
$+0.54$ significantly different from
zero, and it reintroduces a significant negative trend.  Across the three
columns the non-zero mean moves from the recent decade to the full
record, and the trend that the reanalysis removes returns in the
published series at $-0.11$ per decade.  A user of the full historical
record therefore faces a larger mean imbalance in 2025 than in any
earlier vintage.

The trend in the 2025 vintage depends on the window.  On $1959$--$2024$
the reported imbalance has an OLS slope of $-0.11$ per decade
($P=0.02$); on $1959$--$2023$ the slope is $-0.08$ ($P=0.08$), and on
the common $1959$--$2016$ sample it is $-0.07$ ($P=0.19$).
Newey--West standard errors give the same verdict on each window.  The
significance on the full window rests on the $2024$ observation.  We
therefore take the mean, and not the trend, as the statistic that
separates the 2025 vintage from its predecessors.  

\begin{table}[htbp]
  \centering
  \caption{Budget-imbalance mean and trend under three vintages.
    $\overline{\BIM}$ is the sample mean of the published budget
    imbalance in GtC/yr; ``last decade'' is the mean over each
    vintage's final ten years; ``trend'' is the OLS slope per decade.
    Parentheses give $P$-values: for the mean, from a
    heteroskedasticity- and autocorrelation-robust (Newey--West)
    standard error; for the trend, from the OLS standard error,
    reproducing the test reported by \citet{FriedlingsteinNature2026}.  Full
    samples run 1959 to each vintage's last year (2023 for GCB~2024 and
    the consolidated reanalysis, 2024 for GCB~2025).  The consolidated
    row is reconstructed from the deposited correction series, which
    omit the lateral-carbon-export term; its recent-decade mean is
    therefore $-0.10$ rather than the $-0.02$ reported in Table~1 of
    \citet{FriedlingsteinNature2026}.}
  \label{tab:bim_three_vintages}
  \begin{tabular}{llll}
    \toprule
                                                  & Full-sample mean   & Last decade & Trend / decade \\
                                                  & ($1959$--end)      & mean        &                \\
    \midrule
    GCB~2024 \citep{Friedlingstein2024}           & $+0.00$ $(0.99)$   & $-0.44$     & $-0.14$ $(0.003)$ \\
    Reanalysis \citep{FriedlingsteinNature2026}         & $+0.18$ $(0.08)$   & $-0.10$     & $-0.07$ $(0.14)$  \\
    GCB~2025 \citep{GCB2025}           & $+0.54$ $(<0.001)$ & $+0.02$     & $-0.11$ $(0.02)$  \\
    \bottomrule
  \end{tabular}
\end{table}

From here we follow the GCB vintages rather than the reanalysis of \citet{FriedlingsteinNature2026}. We approach the audit in two complementary analyses.  The first is model-free: a three-step procedure that reads the budget-imbalance shift directly off the published GCB data set.  The second uses the dynamic state-space model of \cite{BHK2023}.  A new
vintage appears each year.  Our statistics are defined for any vintage
transition: the audit is a diagnostic to be run on each release rather
than a verdict on one.

Two objects recur throughout the first analysis.  For each sink component the budget reports a single GCB
value, that is, the value the Global Carbon Budget authors have chosen
as the reference. The GCB data set also contains an ensemble of models (about twenty
vegetation models for the land sink and a set of ocean models and
fCO$_2$ products for the ocean sink), whose ensemble mean (the time series of the averages of the individual models) is a natural comparison to the GCB value.  We call the difference between the ensemble mean and the GCB value the \textit{adjustment gap}, and we call the budget imbalance obtained by substituting the ensemble means for the GCB values the \textit{ensemble-mean
imbalance}.  

In the first part of the audit, we analyze the BIM and the adjustment gaps in three steps. The first step 
focuses on the mean of the BIM on the entire sample and on sub-samples and computes these means across the 
GCB vintages 2017 to 2025. The second step locates the origins of the BIM means in the adjustment gaps. 
The third step traces the contributions of additions and removals of ensemble members to the adjustment gaps.
The audit delivers a comprehensive picture of the causes of the BIM mean changes in the ensemble compositions
and in the deviations of the GCB values from the ensemble means.

We attribute the 2025 budget-imbalance mean of $+0.61$~GtC/yr on the common sample 1959--2016 to three sources.  The dominant contribution is
a $+0.40$~GtC/yr adjustment gap on the land sink, the RSS bias correction applied for the first time in the 2025 release.  
The adjustment gap on the ocean sink partially offsets with -0.205 GtC/yr. The remaining
+0.41 GtC/yr is the ensemble-mean imbalance itself, which absorbs the vintage-to-vintage revisions to the underlying inputs (fossil emissions, atmospheric growth, cement carbonation, the
bookkeeping ensemble mean for LUC, including the removal of H\&C2023, which alone contributes +0.195 GtC/yr, and the ensemble means for the
sinks). A separate by-product finding, which the identity does not see because the GCB defines
its land-use emissions as the bookkeeping mean, is a coherent $\sim$0.40 GtC/yr ensemble-wide
downward revision of the DGVM LUC ensemble on nearly all continuing members.

The second part uses the dynamic statistical model of \cite{BHK2023}, hereafter the MDS-GCB
(Multivariate Dynamic Statistical Global Carbon Budget) model,
augmented with three climate-oscillation indices and a stratospheric
aerosol optical depth series, estimated on every GCB vintage
2017--2025 using the common 1959--2016 sample.

The central estimation objects for the model-based audit are the fertilisation slopes $\beta_1$ and $\beta_2$ (the
sensitivities of the two sinks to the atmospheric CO$_2$ stock). For the vintages 2017 to 2024, they are statistically similar.  
The 2025 estimate of $\beta_1$ sits roughly seven averaged standard errors below the eight preceding ones, and the
2025 estimate of $\beta_2$ roughly seven averaged standard errors above. We then show that the introduction of a constant 
in the budget equation partially remedies the imbalance introduced by the 2025 vintage and that, therefore, the RSS bias adjustment
has effectively re-introduced a ``missing sink.'' The model-based analysis also uncovers a dynamic side to the tension in the budget equation that
the model-free part of the audit misses.

The paper is organised as follows.  Section~\ref{sec:audit_panel}
carries out a three-step model-free audit of the GCB ensembles using
only the reported series and descriptive ensemble statistics.  Section~\ref{sec:jrssa_vintages} 
conducts the audit based on the MDS-GCB model of \cite{BHK2023}.
Section~\ref{sec:discussion} discusses what the implications of the findings of the audit for users of the GCB.
Section~\ref{sec:conclusion} concludes.

\FloatBarrier

\section{The model-free audit}
\label{sec:audit_panel}

The Global Carbon Budget closes by construction up to a residual.
The GCB budget imbalance for year $t$ is
\begin{equation}
  \BIM^{\mathrm{GCB}}_t
  = E^{FF}_t + E^{LUC}_t
  - G^{\mathrm{ATM}}_t - S^{LND}_t - S^{OCN}_t
  - S^{\mathrm{CARB}}_t,
  \label{eq:bim_raw}
\end{equation}
where $E^{FF}_t$ is fossil emissions (excluding cement carbonation),
$E^{LUC}_t$ is the GCB land-use change emissions,
$G^{\mathrm{ATM}}_t$ is the atmospheric growth rate, $S^{LND}_t$ and
$S^{OCN}_t$ are the GCB land and ocean sinks, and
$S^{\mathrm{CARB}}_t$ is the cement carbonation sink, reported as a
separate budget term from the GCB 2020 vintage onward.  All quantities are taken from the ``Global Carbon
Budget'' sheet of each vintage's spreadsheet.  The audit is anchored
on the common 1959--2016 sample ($T=58$), the longest sample available
in every vintage from 2017 to 2025.  Robustness checks against each
vintage's own full sample are reported in Appendix~B.

This section presents a three-step model-free audit, run across the
nine GCB vintages 2017--2025 in order to place the 2024--2025
transition against the record.  Step~1 uses the mean of
$\BIM^{\mathrm{GCB}}$ as the trigger and characterises the time
profile of the vintage shift.  Step~2 decompose the BIM mean into the ensemble-mean imbalance and two
sink-specific adjustment gaps (each the difference between a GCB
value and the ensemble mean it is based on).  Step~3 attributes the ensemble-mean shifts
through an Oaxaca-Blinder decomposition \citep{Oaxaca1973,Blinder1973}
to revisions on individual continuing members and contributions from leavers and newcomers.

\subsection{Step 1: BIM mean as audit trigger}
\label{sec:audit_step1}

\begin{figure}[htbp]
  \centering
  \includegraphics[width=\linewidth]{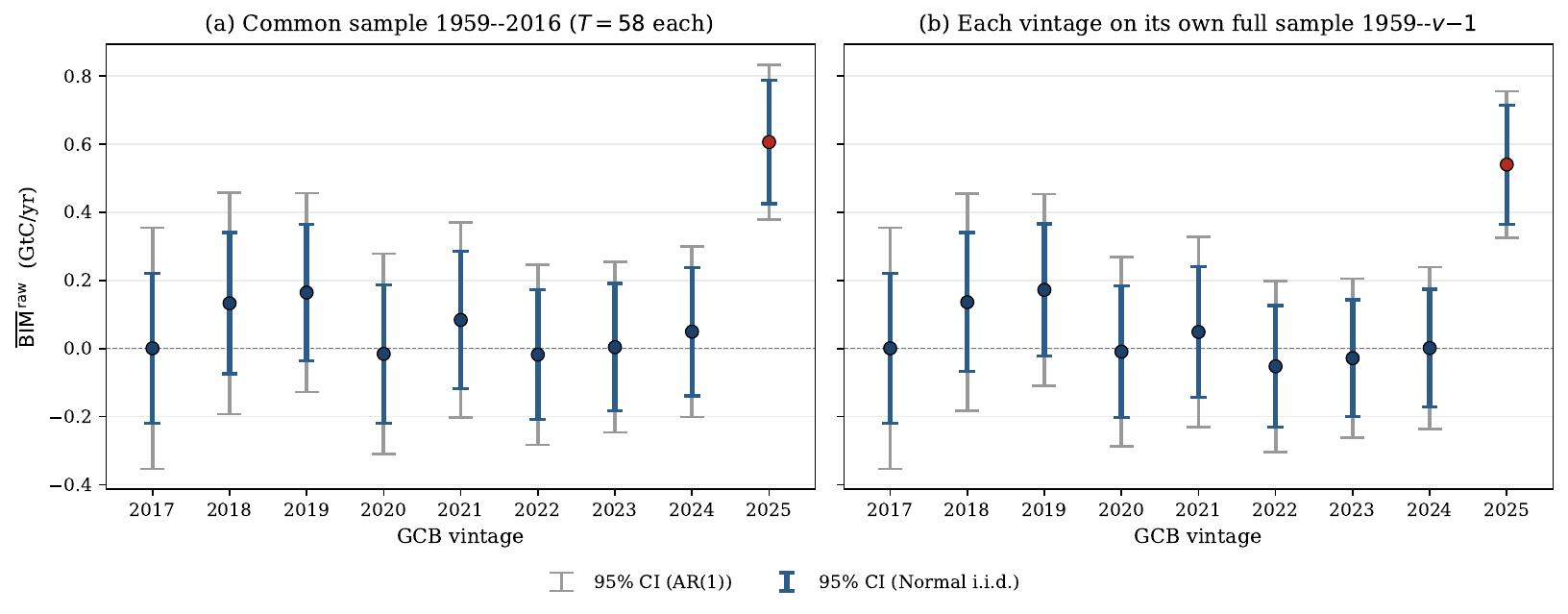}
  \caption{Sample mean of $\BIM^{\mathrm{GCB}}$ per GCB vintage, with
    95\% confidence intervals under an i.i.d.\ Normal assumption
    (narrow blue) and an AR(1)-inflated assumption (wide grey).  The
    2025 vintage is highlighted in red.  Panel~(a) restricts every
    vintage to the common 1959--2016 sample ($T=58$ for each vintage);
    panel~(b) uses each vintage's own full sample 1959 to the year preceding the release.  The
    qualitative finding is identical under either sample: 2025 is the
    only vintage whose confidence interval excludes zero.}
  \label{fig:bim_vintage_means}
\end{figure}

The first-order signal is the sample mean of $\BIM^{\mathrm{GCB}}_t$
on the common 1959--2016 sample.  A well-balanced vintage should have BIM mean near zero; 
a large mean indicates that the published
ensembles and the adjusted GCB values no longer satisfy the
budget identity together.  Figure~\ref{fig:bim_vintage_means} plots
the sample mean of $\BIM^{\mathrm{GCB}}$ per vintage together with two
95\% confidence intervals: one based on an i.i.d.\ Normal assumption and an alternative inflated for first-order
serial correlation. \citet{Bennedsen2021} shows that an AR(1) model adequately captures
the serial dependence in the GCB budget imbalance across vintages.

Eight of the nine vintages have sample means within $\pm 0.17$~GtC/yr
of zero, with confidence intervals comfortably covering zero under
either assumption.  The 2025 vintage is the unique outlier:
$\overline{\BIM}^{\mathrm{GCB}}_{\,2025} = +0.606$~GtC/yr, with both
the Normal CI $[+0.42, +0.79]$ and the AR(1)-inflated CI
$[+0.38, +0.83]$ excluding zero.  The 2024 to 2025 jump of $+0.557$
GtC/yr is more than three times the next-largest consecutive shift
in either direction (2019 to 2020 at $-0.18$).

Table~\ref{tab:bim_subperiods_vintages} reports the mean together with
two loss functions, the root mean-squared error (RMSE) and the mean absolute error (MAE), of $\BIM^{\mathrm{GCB}}$ for
each vintage on its full sample and on successive decades.  2025 stands out 
as the one vintage with a non-zero full-sample mean. The
RMSE and MAE add the size of the imbalance over the entire sample, which the mean does not
measure.  Even where the full-sample mean is near zero, the decade
means swing between roughly $+0.57$ and $-0.44$. Read through the loss functions, the full-sample
imbalance falls steadily across vintages, from an RMSE of $0.85$ (MAE
$0.70$) in 2017 to $0.70$ (MAE $0.57$) in 2023, holds near there in
2024, and then jumps to $0.90$ (MAE $0.73$) in 2025, the largest of
the nine.  The 2025 vintage shifts the mean and raises the RMSE and
MAE above every earlier vintage.  The imbalance is largest in the older
decades (the second-to-last decade mean is $+0.80$) and near zero in the
last decade ($+0.02$).  The corrections apply over the whole record;
their net effect on the imbalance falls outside the last ten years.

The mean and the loss functions answer different questions.  The RMSE
and the MAE measure the size of the annual imbalance, which reflects the
year-to-year uncertainty in all five components and is large in every
vintage.  The mean measures whether the five components close as a set
over the record.  The budget identity constrains the mean and not the
spread, so it is the mean that a user must assume away in order to treat
the published components as mutually consistent.  On the common
$1959$--$2016$ sample the MAE rises from $0.608$ to $0.764$~GtC/yr
between the 2024 and 2025 vintages, while the mean rises from $+0.049$
to $+0.606$.

\begin{table}[htbp]
  \centering
  \scriptsize
  \caption{Budget-imbalance mean, RMSE, and MAE of $\BIM^{\mathrm{GCB}}$ (GtC/yr) for the full sample and by decades for each GCB vintage 2017--2025. The mean is $\overline{\BIM}$, $\mathrm{RMSE}=\sqrt{\overline{\BIM^{2}}}$, and $\mathrm{MAE}=\overline{|\BIM|}$; the loss functions measure the size of the imbalance without the sign cancellation that affects the mean. Each vintage $v$ is evaluated on its full sample $1959$--$(v{-}1)$; decades are counted from the last observation backwards, such that they are not the same across vintages. For example, the last decade of the 2024 vintage is 2014-2023, the last decade of the 2025 vintage is 2015-2024, and so forth. A dash marks a decade a shorter vintage does not reach. The 2025 column is in bold.}
  \label{tab:bim_subperiods_vintages}
  \begin{tabular}{llrrrrrrrrr}
    \toprule
    Vintage $v$ & & 2017 & 2018 & 2019 & 2020 & 2021 & 2022 & 2023 & 2024 & \textbf{2025} \\
    \midrule
    \multirow{3}{*}{Full sample} & mean & $+0.001$ & $+0.136$ & $+0.172$ & $-0.009$ & $+0.048$ & $-0.052$ & $-0.028$ & $+0.001$ & $\mathbf{+0.540}$ \\
     & RMSE & $0.849$ & $0.804$ & $0.779$ & $0.765$ & $0.765$ & $0.718$ & $0.695$ & $0.705$ & $\mathbf{0.899}$ \\
     & MAE & $0.700$ & $0.666$ & $0.636$ & $0.612$ & $0.626$ & $0.602$ & $0.573$ & $0.586$ & $\mathbf{0.730}$ \\
    \addlinespace
    \multirow{3}{*}{Last decade} & mean & $+0.567$ & $+0.507$ & $+0.430$ & $-0.043$ & $-0.275$ & $-0.334$ & $-0.439$ & $-0.443$ & $\mathbf{+0.016}$ \\
     & RMSE & $0.865$ & $0.736$ & $0.676$ & $0.473$ & $0.599$ & $0.554$ & $0.552$ & $0.577$ & $\mathbf{0.622}$ \\
     & MAE & $0.695$ & $0.563$ & $0.519$ & $0.420$ & $0.512$ & $0.478$ & $0.439$ & $0.458$ & $\mathbf{0.431}$ \\
    \addlinespace
    \multirow{3}{*}{\shortstack[l]{Second-to-last\\ decade}} & mean & $-0.406$ & $-0.239$ & $+0.130$ & $+0.033$ & $+0.090$ & $+0.086$ & $+0.043$ & $+0.084$ & $\mathbf{+0.797}$ \\
     & RMSE & $0.750$ & $0.642$ & $0.359$ & $0.524$ & $0.505$ & $0.455$ & $0.495$ & $0.508$ & $\mathbf{0.940}$ \\
     & MAE & $0.530$ & $0.467$ & $0.309$ & $0.374$ & $0.384$ & $0.367$ & $0.398$ & $0.422$ & $\mathbf{0.797}$ \\
    \addlinespace
    \multirow{3}{*}{\shortstack[l]{Third-to-last\\ decade}} & mean & $+0.085$ & $+0.230$ & $+0.254$ & $-0.022$ & $+0.107$ & $-0.105$ & $-0.199$ & $-0.286$ & $\mathbf{+0.270}$ \\
     & RMSE & $1.055$ & $0.999$ & $1.147$ & $1.169$ & $1.021$ & $0.812$ & $0.669$ & $0.673$ & $\mathbf{0.810}$ \\
     & MAE & $0.884$ & $0.823$ & $0.947$ & $0.940$ & $0.775$ & $0.590$ & $0.451$ & $0.518$ & $\mathbf{0.630}$ \\
    \addlinespace
    \multirow{3}{*}{\shortstack[l]{Fourth-to-last\\ decade}} & mean & $-0.425$ & $-0.250$ & $-0.217$ & $-0.384$ & $-0.200$ & $-0.097$ & $+0.007$ & $-0.068$ & $\mathbf{+0.601}$ \\
     & RMSE & $0.881$ & $0.775$ & $0.802$ & $0.826$ & $0.795$ & $0.925$ & $0.986$ & $0.984$ & $\mathbf{1.161}$ \\
     & MAE & $0.800$ & $0.709$ & $0.732$ & $0.692$ & $0.709$ & $0.848$ & $0.891$ & $0.899$ & $\mathbf{0.969}$ \\
    \addlinespace
    \multirow{3}{*}{\shortstack[l]{Fifth-to-last\\ decade}} & mean & $-0.206$ & $-0.127$ & $-0.077$ & $-0.187$ & $-0.248$ & $-0.506$ & $-0.244$ & $-0.074$ & $\mathbf{+0.420}$ \\
     & RMSE & $0.807$ & $0.755$ & $0.681$ & $0.656$ & $0.539$ & $0.711$ & $0.560$ & $0.506$ & $\mathbf{0.628}$ \\
     & MAE & $0.702$ & $0.639$ & $0.547$ & $0.546$ & $0.456$ & $0.628$ & $0.512$ & $0.444$ & $\mathbf{0.532}$ \\
    \addlinespace
    \multirow{3}{*}{\shortstack[l]{Sixth-to-last\\ decade}} & mean & -- & -- & $+0.512$ & $+0.483$ & $+0.652$ & $+0.376$ & $+0.288$ & $+0.290$ & $\mathbf{+0.685}$ \\
     & RMSE & -- & -- & $0.799$ & $0.742$ & $0.947$ & $0.709$ & $0.691$ & $0.657$ & $\mathbf{0.814}$ \\
     & MAE & -- & -- & $0.761$ & $0.698$ & $0.872$ & $0.636$ & $0.615$ & $0.558$ & $\mathbf{0.688}$ \\
    \addlinespace
    \bottomrule
  \end{tabular}
\end{table}

Before proceeding to the decomposition, we illustrate the effects of the 2025 vintage on two derived
budget statistics, the airborne fraction and the sink rate.  Define net anthropogenic emissions
\begin{equation}
  E_t \;=\; E^{FF}_t + E^{LUC}_t - S^{\mathrm{CARB}}_t,
  \label{eq:panel_E_net}
\end{equation}
so that the cement-carbonation sink is subsumed in the anthropogenic emissions term
and the budget identity simplifies to
$\BIM^{\mathrm{GCB}}_t = E_t - G^{ATM}_t - S^{LND}_t - S^{OCN}_t$.
Define
\begin{equation}
  \mathrm{AF}_t \;=\; \frac{G^{ATM}_t}{E_t},
  \qquad
  \mathrm{SR}^{\,1}_t \;=\; \frac{S^{LND}_t + S^{OCN}_t}{C_t - C_{1750}},
  \qquad
  \mathrm{SR}^{\,2}_t \;=\; \frac{E_t - G^{ATM}_t}{C_t - C_{1750}},
  \label{eq:panel_af_sr}
\end{equation}
with $C_t = C_{1959} + \sum_{s\le t} G^{ATM}_s$ and $C_{1750}$ the
pre-industrial concentration level.  The two sink-rate definitions differ
in the numerator: $\mathrm{SR}^{\,1}$ uses the published land and
ocean sinks directly; $\mathrm{SR}^{\,2}$ uses the budget identity to
back out an implied total sink from net emissions minus atmospheric
growth.  Subtracting,
\begin{equation*}
  \mathrm{SR}^{\,2}_t - \mathrm{SR}^{\,1}_t
  \;=\; \frac{\BIM^{\mathrm{GCB}}_t}{C_t - C_{1750}}.
\end{equation*}
The two definitions agree when the published imbalance is near zero;
they diverge in proportion to the budget imbalance.

\begin{figure}[htbp]
  \centering
  \includegraphics[width=\linewidth]{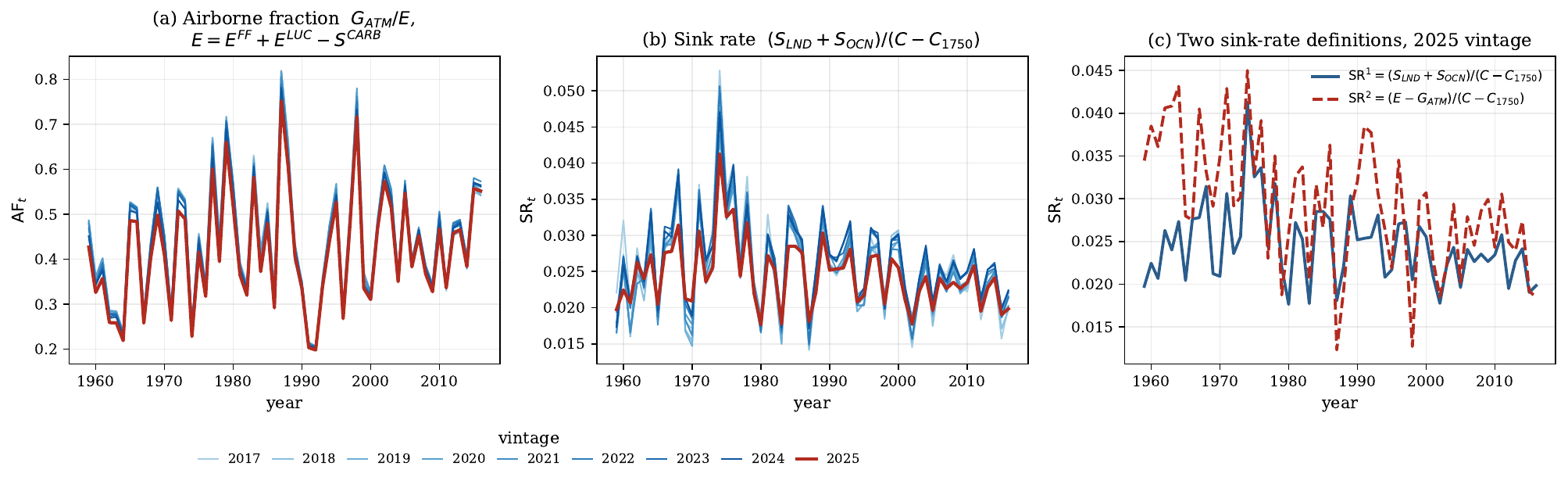}
  \caption{Airborne fraction and sink
    rate, computed directly from the published GCB series with net
    anthropogenic emissions $E_t = E^{FF}_t + E^{LUC}_t -
    S^{\mathrm{CARB}}_t$.  Panel~(a) plots
    $\mathrm{AF}_t = G^{ATM}_t / E_t$ for each GCB vintage 2017--2025
    on the common 1959--2016 sample.  Panel~(b) plots
    $\mathrm{SR}^{\,1}_t = (S^{LND}_t + S^{OCN}_t) / (C_t - C_{1750})$
    on the same sample.  Vintages 2017--2024 are drawn in graduated
    blue; 2025 in red.  Panel~(c) compares $\mathrm{SR}^{\,1}_t$
    against the budget-identity-implied
    $\mathrm{SR}^{\,2}_t = (E_t - G^{ATM}_t) / (C_t - C_{1750})$
    for the 2025 vintage only.  The gap between the two definitions
    equals $\BIM^{\mathrm{GCB}}_t / (C_t - C_{1750})$ exactly and
    traces the budget imbalance noted in Step~1.}
  \label{fig:panel_af_sr_vintages}
\end{figure}

In Figure~\ref{fig:panel_af_sr_vintages}, the airborne fraction in
panel~(a) looks stable across the nine vintages: the large
year-to-year features (the El Niño peaks, the post-Pinatubo dip) are
common to all vintages and dominate the panel, so the 2025 path
overlaps the eight-vintage envelope for most of the sample.  The
sink-rate panel (b) shows a different pattern.  The 2025 series lies
below the other eight vintages across most of the sample.  The downward shift is the direct
consequence of the methodological adjustments to $S^{LND}$ and
$S^{OCN}$ documented in the GCB 2025 release: the RSS bias correction
reduces the published land sink, and the correction does not have a
fully offsetting counterpart on the ocean side, so the numerator of
$\mathrm{SR}^{\,1}$ falls relative to the eight-vintage cluster.

The airborne-fraction means lie between $0.434$ and
$0.446$ for the 2017--2024 vintages and drop to $0.419$ in 2025; the
sink-rate means run from $0.0249$ to $0.0266$ and drop to $0.0246$.
In relative terms both fall by about five per cent in 2025, so the
level shift that the sink-rate panel makes visible is present in the
airborne fraction as well; the airborne fraction shows it less
visibly, because its larger year-to-year variation dominates the panel.

Panel (c) isolates the consequence directly.  Algebraically, the two sink-rate definitions
$\mathrm{SR}^{\,1}$ and $\mathrm{SR}^{\,2}$ should be identical: the
carbon-budget identity, with $S^{\mathrm{CARB}}$ subsumed in $E$,
links $E_t - G^{ATM}_t$ to $S^{LND}_t + S^{OCN}_t$, so the two
numerators differ only by $\BIM^{\mathrm{GCB}}_t$.  In any vintage
with a near-zero published imbalance, the two curves coincide to
within numerical noise.  In the 2025 vintage they do not.  The gap
$\mathrm{SR}^{\,2}_t - \mathrm{SR}^{\,1}_t$ equals
$\BIM^{\mathrm{GCB}}_t / (C_t - C_{1750})$ and tracks the budget
imbalance throughout the sample, averaging about
$0.57 / 200 \approx 0.003$ at recent values of
$C_t - C_{1750} \approx 200$~GtC.  A consequence of the non-zero
imbalance mean in 2025 is therefore that the sink rate, as a
statistic derived from the published GCB series, becomes ill-defined:
depending on which arithmetic the reader uses, the published numbers
return one of two answers that no longer coincide.

\subsection{Step 2: The contribution of the adjustment gaps to the BIM means}
\label{sec:audit_step2}

For each component $X \in \{\mathrm{LUC}, S_{\mathrm{LND}},
S_{\mathrm{OCN}}\}$ define the adjustment gap as the difference between the
ensemble mean $\bar X$ (the average of the individual models published in the
GCB data set) and the GCB value $X^{\mathrm{GCB\text{-}pref}}$ (the value the Global Carbon Budget
authors have chosen as the reference),
\begin{equation}
  \mathrm{gap}_X = \bar X^{\mathrm{ens}} - X^{\mathrm{GCB\text{-}pref}}.
  \label{eq:gap_def}
\end{equation}
For LUC, the ensemble is the bookkeeping models (BLUE, OSCAR,
LUCE, and H\&C2023 in 2024); for $S_{\mathrm{LND}}$, it is the DGVM
ensemble; for $S_{\mathrm{OCN}}$, it is the combined GOBM and fCO$_2$
product ensemble.  Define the ensemble-mean imbalance as the
budget imbalance that results from using the ensemble means instead of
the GCB values, that is, by replacing the GCB sink and LUC values with their
ensemble means in \eqref{eq:bim_raw},
\begin{equation*}
  \BIM^{\mathrm{ens}}_t
    = E^{FF}_t + \bar E^{LUC,\mathrm{ens}}_t
    - G^{\mathrm{ATM}}_t - \bar S^{LND,\mathrm{ens}}_t
    - \bar S^{OCN,\mathrm{ens}}_t - S^{\mathrm{CARB}}_t.
\end{equation*}
Subtracting $\BIM^{\mathrm{ens}}$ from $\BIM^{\mathrm{GCB}}$ gives
the exact algebraic identity
\begin{equation}
  \BIM^{\mathrm{GCB}}
  \;=\; \BIM^{\mathrm{ens}}
        - \mathrm{gap}_{\mathrm{LUC}}
        + \mathrm{gap}_{S_{\mathrm{LND}}}
        + \mathrm{gap}_{S_{\mathrm{OCN}}}.
  \label{eq:audit_decomp}
\end{equation}
The GCB land-use emissions $E^{LUC}$ are reported as the
arithmetic mean of the bookkeeping (BK) ensemble, year by year, with no
adjustment.  The BK ensemble mean therefore equals the GCB
$E^{LUC}$ exactly in both vintages, and $\mathrm{gap}_{\mathrm{LUC}}
\equiv 0$ identically.  Equation~\eqref{eq:audit_decomp} then
reduces to
\begin{equation}
  \BIM^{\mathrm{GCB}}
  \;=\; \BIM^{\mathrm{ens}}
        + \mathrm{gap}_{S_{\mathrm{LND}}}
        + \mathrm{gap}_{S_{\mathrm{OCN}}}.
  \label{eq:audit_decomp_red}
\end{equation}
Each of the three terms has an audit interpretation.
$\BIM^{\mathrm{ens}}$ is the ensemble-mean imbalance: the budget
imbalance the underlying model ensembles would have produced if the
GCB had used ensemble means for the sinks.
$\mathrm{gap}_{S_{\mathrm{LND}}}$ and $\mathrm{gap}_{S_{\mathrm{OCN}}}$
are the cumulative effects of the GCB sink adjustments away from the
ensemble means.

\begin{table}[h]
  \centering
  \caption{Step 2 identity decomposition across vintages. Sample means in GtC/yr on the common $1959$--$2016$ sample ($T=58$). Row identity: $\BIM^{\mathrm{GCB}} = \BIM^{\mathrm{ens}} + \mathrm{gap}_{S_{\mathrm{LND}}} + \mathrm{gap}_{S_{\mathrm{OCN}}}$.}
  \label{tab:audit_step2_vintages}
  \begin{tabular}{lrrrr}
    \toprule
    Vintage & $\BIM^{\mathrm{GCB}}$ & $\BIM^{\mathrm{ens}}$ & $\mathrm{gap}_{S_{\mathrm{LND}}}$ & $\mathrm{gap}_{S_{\mathrm{OCN}}}$ \\
    \midrule
    2017 & $+0.001$ & $+0.045$ & $+0.000$ & $-0.045$ \\
    2018 & $+0.133$ & $+0.192$ & $-0.000$ & $-0.059$ \\
    2019 & $+0.164$ & $+0.221$ & $-0.005$ & $-0.057$ \\
    2020 & $-0.016$ & $+0.047$ & $-0.000$ & $-0.062$ \\
    2021 & $+0.084$ & $+0.133$ & $+0.000$ & $-0.049$ \\
    2022 & $-0.018$ & $+0.042$ & $-0.000$ & $-0.060$ \\
    2023 & $+0.004$ & $+0.070$ & $+0.000$ & $-0.066$ \\
    2024 & $+0.049$ & $+0.111$ & $+0.000$ & $-0.062$ \\
    \bfseries{}2025 & $\mathbf{+0.606}$ & $\mathbf{+0.411}$ & $\mathbf{+0.400}$ & $\mathbf{-0.205}$ \\
    \bottomrule
  \end{tabular}
\end{table}

The sample-mean decomposition, run across all nine vintages on the
common 1959--2016 sample, is reported in
Table~\ref{tab:audit_step2_vintages}. Two patterns are visible in the table.  First, the land-sink gap
$\mathrm{gap}_{S_{\mathrm{LND}}}$ is essentially zero in every
vintage except 2025 (a small $-0.005$~GtC/yr residual in 2019 is
attributable to rounding in the published BIM), where it jumps to
$+0.400$~GtC/yr.  Until 2025 the GCB land sink coincided with the DGVM
Multi-model mean year by year; the 2025 release introduced a Replaced
Sinks and Sources (RSS) bias correction \citep[Sect.~2.6.1
of][]{GCB2025}, applied for the first time in this vintage,
which the spreadsheet reports as a distinction between the ``GCB''
column (adjusted) and a new ``GCB unadjusted'' column equal to the
DGVM ensemble mean.  This RSS adjustment is the single largest
contributor to the 2025 budget imbalance.

Second, the ocean gap ($\mathrm{gap}_{S_{\mathrm{OCN}}}$) sits in a
tight band between $-0.045$ and $-0.066$~GtC/yr across vintages
2017--2024, with year-to-year reversals, and then increases to
$-0.205$~GtC/yr in 2025, roughly three times the magnitude of any
earlier vintage.  The residual
$\BIM^{\mathrm{ens}}$ ranges from $+0.042$ to $+0.221$~GtC/yr
across vintages 2017--2024 and rises to $+0.411$~GtC/yr in 2025,
nearly twice the largest pre-2025 value.  These movements reflect
vintage-to-vintage
revisions to fossil emissions $E^{FF}_t$, atmospheric growth
$G^{\mathrm{ATM}}_t$, the cement carbonation sink
$S^{\mathrm{CARB}}_t$ (from 2020 onward), the bookkeeping LUC
ensemble mean, and the ensemble means for the sinks themselves.  We
return to the drivers of $\overline{\BIM}^{\mathrm{ens}}$ in
Step~3.

In 2025 the identity~\eqref{eq:audit_decomp_red} reads, on the common
sample, $\overline{\BIM}^{\mathrm{GCB}} = 0.411 + 0.400 - 0.205 =
+0.606$~GtC/yr: the ensemble-mean imbalance, the land-sink gap, and the
ocean-sink gap.  The land-sink gap is the dominant term, two-thirds of
the total and new in this vintage; the ocean-sink gap partially offsets
it; and the ensemble-mean imbalance carries the remainder. 

The identity in \eqref{eq:audit_decomp_red} does not involve the LUC
side: the GCB defines its published $E^{LUC}$ as the
bookkeeping-ensemble mean, so $\mathrm{gap}_{\mathrm{LUC}} \equiv 0$ by
construction.  An independent check therefore uses the second LUC
ensemble, the DGVM LUC models.  For vintages 2017--2024 the DGVM LUC ensemble mean sits
within $\pm 0.18$~GtC/yr of the bookkeeping ensemble mean, most
often slightly above.  In 2025 the two cross and separate by
$+0.55$~GtC/yr: the BK mean rose from $+1.487$ to $+1.748$~GtC/yr on
the common sample, while the DGVM LUC mean fell from $+1.593$ to
$+1.196$.  The GCB chooses the bookkeeping models (BK) as its $E^{LUC}$ reference in every
vintage; Step~3 attributes the BK shift almost entirely to the removal
of one ensemble member. The DGVM LUC shift to a coherent
ensemble-wide revision on continuing members does not enter the budget equation.

Both sink gaps change substantially in 2025 rather than drift across vintages.  The
land-sink gap holds at zero through 2017--2024 and jumps to
$+0.400$~GtC/yr; the ocean gap holds a narrow $-0.045$ to
$-0.066$~GtC/yr band and then increases to $-0.205$~GtC/yr.  Both are
consistent with the 2025 release introducing the corrections in a
single methodological update.  We return to the residual
$\overline{\BIM}^{\mathrm{ens}}$ and to the ensemble-mean shifts in
the discussion (Section~\ref{sec:discussion}).

\subsection{Step 3: Per-member attribution of ensemble-mean shifts}
\label{sec:audit_step4}

The ensemble means themselves shift between vintages; Step~3
attributes each shift to specific members of the underlying ensemble
through an Oaxaca-Blinder decomposition.  Let
$C$ index the continuing members present in both vintages, ``out''
the leavers (present in 2024 but not 2025), and ``in'' the newcomers
(present in 2025 but not 2024).  Let $m_C(t)$ denote the mean over
the continuing members at vintage $t$, $m_{\mathrm{out}}$ the mean
of the leavers' 2024 values, $m_{\mathrm{in}}$ the mean of the
newcomers' 2025 values, and $N_t$ the total ensemble size at vintage
$t$.  Then, for each $X \in \{\mathrm{LUC}, S_{\mathrm{LND}},
S_{\mathrm{OCN}}\}$, the ensemble-mean change decomposes exactly as
\begin{equation}
  \Delta \bar X \;=\; \underbrace{m_C(2025) - m_C(2024)}_{\Delta_{\mathrm{rev}}}
  \;+\; \underbrace{\tfrac{N_{\mathrm{in}}}{N_{2025}}
        \bigl(m_{\mathrm{in}} - m_C(2025)\bigr)}_{\Delta_{\mathrm{in}}}
  \;-\; \underbrace{\tfrac{N_{\mathrm{out}}}{N_{2024}}
        \bigl(m_{\mathrm{out}} - m_C(2024)\bigr)}_{\Delta_{\mathrm{out}}}.
  \label{eq:oaxaca}
\end{equation}
Continuing-member $i$ contributes $(x_{i,2025} - x_{i,2024}) / N_C$
to $\Delta_{\mathrm{rev}}$, each leaver $i$ contributes
$(m_C(2024) - x_{i,2024}) / N_{2024}$ to $-\Delta_{\mathrm{out}}$, and
each newcomer $i$ contributes $(x_{i,2025} - m_C(2025)) / N_{2025}$
to $\Delta_{\mathrm{in}}$.  Member contributions are reported in raw
GtC/yr ensemble-mean units.

\begin{table}[h]
  \centering
  \caption{Step 3 Oaxaca-Blinder decomposition of the 2024 to 2025
    transition, in GtC/yr.  $\Delta \bar X$ is the ensemble-mean shift,
    decomposed via \eqref{eq:oaxaca} into revision on continuing
    members ($\Delta_{\mathrm{rev}}$), inflow from newcomers
    ($\Delta_{\mathrm{in}}$), and outflow from leavers
    ($-\Delta_{\mathrm{out}}$).  Computed on the common
    $1959$--$2016$ sample; sub-row residuals from NaN-handling at the
    member level are below $0.01$~GtC/yr.}
  \label{tab:audit_step4}
  \begin{tabular}{lrrrr}
    \toprule
    Ensemble                       & $\Delta \bar X$ & $\Delta_{\mathrm{rev}}$ & $\Delta_{\mathrm{in}}$ & $-\Delta_{\mathrm{out}}$ \\
    \midrule
    BK (LUC ensemble)           & $+0.261$ & $+0.066$ & $\phantom{+}0.000$ & $+0.195$ \\
    DGVM LUC                    & $-0.397$ & $-0.380$ & $-0.024$ & $\phantom{+}0.000$ \\
    DGVM $S_{\mathrm{LND}}$     & $-0.077$ & $-0.004$ & $-0.115$ & $+0.043$ \\
    Ocean                       & $+0.015$ & $+0.015$ & $\phantom{+}0.000$ & $\phantom{+}0.000$ \\
    \bottomrule
  \end{tabular}
\end{table}

Table~\ref{tab:audit_step4} summarizes the decomposition: For the bookkeeping ensemble, the $-\Delta_{\mathrm{out}}$ column carries
most of the $\Delta \bar X$: the removal of the H\&C2023 bookkeeping
model, whose 1959--2016 mean was $+0.903$~GtC/yr (well below the
three continuing members at $+1.682$ on average), contributes
$+0.195$~GtC/yr of the $+0.261$ shift in the bookkeeping ensemble mean.  Revision on
the three continuing members contributes only $+0.066$~GtC/yr.  No
members were added.  The shift in the bookkeeping ensemble mean is a single-member composition
effect.

For the DGVM LUC ensemble, the dominant component is the revision
column $\Delta_{\mathrm{rev}} = -0.380$~GtC/yr.  Of the twenty
continuing DGVMs (treating the JULES and JULES-ES entries as the same
model), the
largest individual revisions are VISIT ($1.65 \to 0.10$~GtC/yr), OCN
($1.96 \to 0.53$), ISAM ($1.97 \to 0.95$), CLM6.0 ($2.62 \to 1.77$),
ELM ($2.10 \to 1.46$), LPJ-GUESS ($1.66 \to 1.07$), ED
($2.55 \to 1.98$), JSBACH ($1.62 \to 1.09$), CLASSIC
($1.27 \to 0.85$), and CABLE-POP ($1.80 \to 1.37$).  Nearly every
continuing member lost LUC emissions between vintages; this is a
coherent ensemble-wide shift, not a small composition
effect.  Newcomers contribute a further $-0.024$~GtC/yr (driven by
ELM-FATES, $+0.61$~GtC/yr 1959--2016 mean, with CLM-FATES at
$+1.46$); no DGVM LUC members were removed.

For the DGVM land-sink ensemble, the ensemble mean barely changes
($\Delta \bar X = -0.077$~GtC/yr) and revision on continuing members
is essentially zero ($\Delta_{\mathrm{rev}} = -0.004$).  The
composition-in component is $\Delta_{\mathrm{in}} = -0.115$, driven
by the new ELM-FATES ($1.05$~GtC/yr 1959--2016 mean) and VISIT-UT
($1.59$) members sitting well below the continuing-cohort mean of
$2.209$.  This shift,
however, is small relative to the $+0.400$~GtC/yr gap that Step~2
isolated on the same ensemble.  The bulk of
$\mathrm{gap}_{S_{\mathrm{LND}}}$ therefore comes from the
GCB adjustment and not
from movement in the underlying DGVM ensemble itself.

For the ocean ensemble, there are no composition changes between 2024
and 2025 once the MOM6-COBALT (Princeton) and Princeton entries are
matched as the same product: all 19 data products are present in both
vintages.
The ensemble-mean shift of $+0.015$~GtC/yr is therefore entirely a
revision on continuing members.  The two largest revisions are
Princeton ($+1.47 \to +1.88$~GtC/yr 1959--2016 mean) and
UoEX-UEPFFNU (formerly Watson, $+2.98 \to +2.64$), which run in
opposite directions, with JMA-MLR ($+2.86 \to +3.05$) and ACCESS
($+2.27 \to +2.09$) contributing further partially offsetting
movements.  The net ensemble-mean shift, $+0.015$~GtC/yr, is much
smaller than the $-0.205$~GtC/yr ocean gap that Step~2 records for
2025.  As on the land side, the ocean BIM signal
comes from a GCB adjustment away from the ensemble rather
than from ensemble-level movement.

We have applied the same Oaxaca-Blinder decomposition to each
consecutive vintage transition 2017--2018 through 2023--2024.
Appendix~B reports the full set of seven decompositions, one per ensemble
per transition.  No transition before 2024--2025 produces an ensemble-mean
shift larger than $\pm 0.32$~GtC/yr (the largest is the ocean ensemble
2020--2021 at $+0.310$~GtC/yr, driven by a composition reshuffle of
five new and three removed members), and no single decomposition component
exceeds $\pm 0.29$~GtC/yr.  The 2024--2025 transition is the only one
whose dominant component (DGVM LUC, $\Delta_{\mathrm{rev}}=-0.380$)
exceeds the largest comparable component in any earlier transition by
more than a tenth of a GtC/yr.

\subsection{Counterfactual: H\&C2023 retained}
\label{sec:audit_step5}

The bookkeeping ensemble member that was removed is close to a constant
offset, which raises the question of how much of the 2025 shift it
accounts for on its own.  The removal can be undone arithmetically.
H\&C2023 covers $1959$--$2016$ in full, so on the common sample the
counterfactual requires no extrapolation.  We hold the model at the
values published in the 2024 vintage, add it back as a fourth member of
the 2025 bookkeeping ensemble, recompute the ensemble mean, and pass the
difference through the budget identity.  Every other component stays at
its 2025 value.

Holding the model at its 2024 values costs little.  Between its last two
appearances its common-sample series was revised by $-0.011$~GtC/yr in
the mean, with a standard deviation of $0.021$ and a largest single-year
difference of $0.069$~GtC/yr.

With H\&C2023 restored, the bookkeeping ensemble mean falls from
$+1.748$ to $+1.537$~GtC/yr and the budget imbalance falls from $+0.606$
to $+0.395$~GtC/yr.\footnote{The Oaxaca-Blinder leaver term in
Table~\ref{tab:audit_step4} is $+0.195$~GtC/yr rather than $+0.211$.  The
two answer different questions.  The decomposition measures the
removal's contribution to the change in the ensemble mean between the
two vintages, with the continuing members at their 2024 values; the
counterfactual asks what the 2025 ensemble mean would be with H\&C2023
added to the members at their 2025 values.}  The $95\%$ confidence
interval for the counterfactual mean is $[+0.21, +0.58]$ under the
i.i.d.\ Normal assumption and $[+0.17, +0.62]$ under the AR(1)
inflation, and excludes zero in both cases.

The removal of H\&C2023 therefore accounts for about a third of the 2025
imbalance.  The remainder is larger than the full-sample imbalance mean
of any vintage from 2017 to 2024.

\subsection{Findings for the 2024 to 2025 transition}
\label{sec:audit_findings}

The audit produces four concrete findings for the 2024 to 2025 transition, in order of their bearing on the budget imbalance. First, the GCB 2025 release applies the RSS bias correction to the land sink, $-0.400$~GtC/yr on the common sample. In the 2024 vintage the published (``GCB'') land sink equals the DGVM ensemble mean exactly. The 2025 vintage adds a new ``GCB unadjusted'' column, still equal to the DGVM ensemble mean, while its ``GCB'' column now lies $0.400$~GtC/yr below it (averaged over the common 1959--2016 window used in the audit). The difference is the RSS reduction. This is the largest single adjustment the GCB documents.

Second, a $-0.205$~GtC/yr adjustment gap on the ocean sink that partially offsets the RSS reduction. Two corrections in the 2025 release raise the published ocean sink above the GOBM and fCO$_2$ ensemble mean: a temperature correction to the fCO$_2$ products and a scaling of the GOBM ensemble mean. The adjustment gap for the ocean deepens from a $-0.05$ to $-0.07$~GtC/yr band to $-0.205$~GtC/yr. Like the land-sink adjustment gap, it is a GCB adjustment away from the ensemble, not a movement of the ensemble itself.

Third, the ensemble-mean imbalance $\overline{\BIM}^{\mathrm{ens}}$ stands at $+0.411$~GtC/yr, independent of the adjustment gaps. It absorbs the vintage-to-vintage revisions to the underlying inputs, chiefly the removal of H\&C2023 from the bookkeeping ensemble ($+0.195$~GtC/yr through the published $E^{LUC}$) and the DGVM land-sink newcomers ($-0.115$~GtC/yr on the DGVM land-sink mean), with the fossil, atmospheric-growth, and cement revisions netting to about $-0.01$.

Fourth, and with no effect on the budget imbalance, the DGVM LUC ensemble underwent a coherent $-0.380$~GtC/yr downward revision across nearly all continuing members. Because the GCB takes its $E^{LUC}$ from the bookkeeping mean and not the DGVM LUC mean, this revision never enters the BIM; Step~3 reports it only as an independent observation.

\FloatBarrier

\section{The model-based audit}
\label{sec:jrssa_vintages}

\subsection{Structural parameters}
\label{sec:jrssa_vintages_struct}

We estimate the model of \cite{BHK2023} (described in Appendix~\ref{sec:jrssa_brief}) on every GCB
vintage from 2017 to 2025, using the common 1959--2016 sample
($T = 58$) for direct comparability across vintages.  The persistent budget imbalance documented in Section~\ref{sec:audit_panel}
moves the fertilisation slopes $\beta_1$ and $\beta_2$ that capture the linear dependence of the land and ocean sinks, respectively, on concentrations. Over the common 1959--2016 sample, concentrations are dominated by a smooth upward trend. An imbalance that averages to zero merely adds estimation noise to the estimates of $\beta_1$ and $\beta_2$. A trending imbalance, however,
injects a trend-like component into the budget equation, and since concentrations are the only strongly trending regressor in the sink
equation, the estimator can absorb that component only through the fertilisation slope. Whatever share of the imbalance is allocated
to a sink is read as additional sink response to rising CO$_2$. 

\begin{figure}[htbp]
  \centering
  \includegraphics[width=\linewidth]{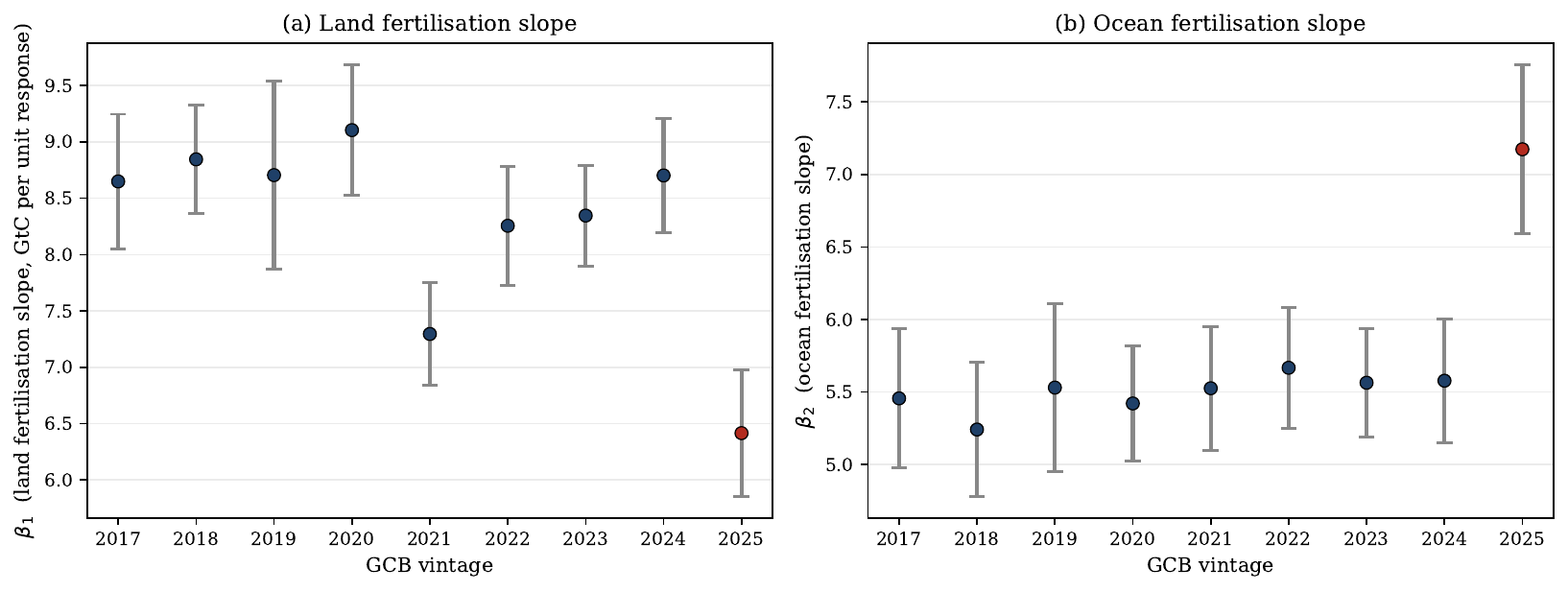}
  \caption{Land and ocean fertilisation slopes $\beta_1$ and $\beta_2$
    of the MDS-GCB model, estimated on each GCB vintage 2017--2025
    using the common 1959--2016 sample.  Bars are $\pm 1.96$
    standard errors.  Vintages 2017--2024 are in blue; the 2025 vintage is
    highlighted in red.}
  \label{fig:jrssa_beta_vintages}
\end{figure}

Figure~\ref{fig:jrssa_beta_vintages} shows the estimated land and ocean fertilisation slopes
$\beta_1, \beta_2$ across the vintages. The vintage-to-vintage movements in $\hat\beta_1$ and $\hat\beta_2$ are the model-based shadow of the imbalance revisions traced model-free in Section~\ref{sec:audit_panel}. The budget imbalance drift added in the 2025 vintage passes into the estimated slopes.

The qualitative pattern is the same as for the model-free audit in
Section~\ref{sec:audit_panel}: eight vintages cluster tightly, with the exception of 2021 for $\beta_1$, and the
2025 vintage sits outside that cluster on both panels.  The land
fertilisation slope $\beta_1$ for vintages 2017--2024 ranges from
$+7.30$ to $+9.10$, with vintage-by-vintage standard errors between
$0.23$ and $0.43$; the 2025
estimate is
$+6.42\,(0.29)$, roughly seven averaged standard errors below the
2017--2024 mean.  The ocean fertilisation slope $\beta_2$ for
2017--2024 ranges from $+5.24$ to $+5.67$, with standard errors
$0.19$--$0.30$; the
2025 estimate is $+7.17\,(0.30)$, roughly seven averaged standard
errors above the 2017--2024 mean.  The two slopes move in opposite
directions in 2025 and the magnitudes are broadly similar.

The full 20-parameter table for all vintages is collected in
Appendix~\ref{app:full_parameters}. We note here that the main instabilities
introduced by the 2025 vintage occur, apart from the fertilisation coefficients, in
the intercepts of the sink equations, which have to move together with the 
fertilisation coefficients, and in the serial correlation parameters 
of the concentrations equation and of the ocean sink equation. 
We return to the issue of serial correlation in the measurement errors below.

\begin{figure}[htbp]
  \centering
  \includegraphics[width=\linewidth]{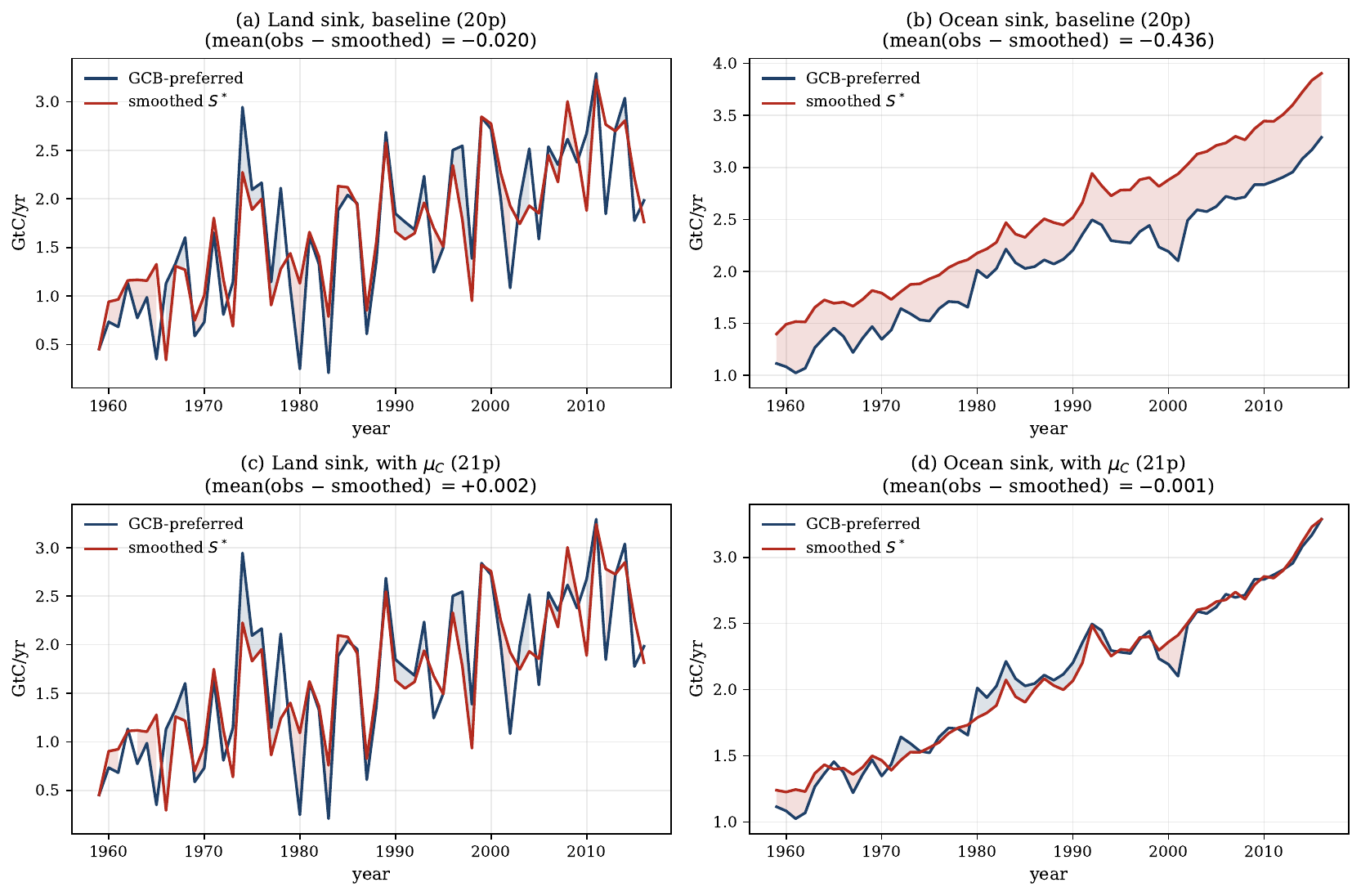}
  \caption{Smoothed land and ocean sink states against the
    published GCB series on the 2025 vintage, common
    $1959$--$2016$ sample.  Top row: baseline MDS-GCB ($20$
    parameters), $\hat\phi_3 = 0.99$.  Bottom row:
    $\mu_C$-augmented MDS-GCB ($21$ parameters), $\hat\phi_3 = 0.61$.
    The baseline fit lifts $S^{OCN*}$ about $0.44$~GtC/yr
    above the published ocean series across the whole sample,
    absorbing the $+0.61$~GtC/yr budget imbalance through the ocean
    AR(1) measurement residual at the upper limit of its persistence.  Adding
    a constant $\mu_C$ to the budget-identity state equation routes
    the imbalance into $\hat\mu_C = -0.59$~GtC/yr instead, leaving
    the smoothed sinks to track their published series.}
  \label{fig:smoothed_sinks_2025}
\end{figure}

\subsection{Drift in the budget equation}
\label{sec:audit_model_bim}

The MDS-GCB model imposes the carbon-budget identity exactly
as a deterministic state equation on $C^{*}$.  When the published
vintage carries a persistent budget imbalance, as the 2025 vintage
does with $\overline{\BIM}^{\mathrm{GCB}}_{\,2025} = +0.61$ GtC/yr
on $1959$--$2016$, the budget equation has no residual to absorb the imbalance. 
The fit reconciles it by deforming the latent states,
and it does so through the two measurement residuals with enough
memory to hold a long-lived offset.  On the sink side, the ocean
measurement residual $X^{(3)}$, an AR(1) process, is driven to its
upper limit ($\hat\phi_3 = 0.99$), an effective random walk, and the
smoothed $S^{OCN*}$ ends up sitting about $+0.44$ GtC/yr above the
published GOBM and fCO$_2$ ensemble across the whole sample (the
smoothed $S^{LND*}$ tracks the published $S^{LND}$ to within $0.02$
GtC/yr).  On the cumulative-CO$_2$ side, the persistence $\phi_1$ on
the measurement residual $X^{(1)}$ is driven to its upper limit
$+1.00$ in the same fit.  The top row of
Figure~\ref{fig:smoothed_sinks_2025} plots the two smoothed-vs-published
sink series on the 2025 vintage and makes the ocean inflation apparent.

The offset is not identified from the four series in the model.  Nothing
in the land, ocean, emissions, or concentration data distinguishes an
ocean sink $0.44$ GtC/yr above the published ensemble from a budget that
does not close, and the deterministic identity routes the discrepancy
into the one residual with the persistence to hold it.
\citet{GCB2025} names an underestimated ocean sink as one
possible source of the imbalance (Sect.~3.9.2), on evidence outside the
four series we use.  We do not test that explanation.  We locate the
discrepancy in the fit; we do not identify its cause.

The obvious remedy is to give the budget identity a level parameter as follows.
Define the augmented state equation
\begin{equation}
  C^{*}_{t+1} - C^{*}_{t}
   \;=\; X^{(E)}_t + E^{*}_t + d
       - S^{LND*}_{t+1} - S^{OCN*}_{t+1} + \mu_C,
  \label{eq:jrssa_muC}
\end{equation}
with $\mu_C$ an unrestricted free parameter and the rest of the model
unchanged.  Estimated on the 2025 vintage on the same $1959$--$2016$
sample, $\hat\mu_C = -0.59$ (0.09) GtC/yr, $\phi_3$ returns to $0.61$
(within the $2017$--$2024$ range), the smoothed $S^{OCN*}$ tracks
the published ocean series to within $0.001$ GtC/yr, and the
likelihood improves by $11.0$ log-units against the baseline
(likelihood-ratio statistic $22.08$ on one degree of freedom). 

\begin{table}[h]
  \centering
  \footnotesize
  \caption{Comparison of the baseline MDS-GCB model and the
    $\mu_C$-augmented variant on the 2025 vintage, common $1959$--$2016$
    sample. Standard errors in parentheses are computed by the delta
    method from the regularised inverse Hessian; standard errors for
    $c_1, c_2$ are the diagonal of the smoothed-state covariance at
    the last observation.
    The augmentation adds one parameter, an unconstrained constant $\mu_C$
    on the carbon-budget state equation, and is estimated by maximum
    likelihood. The likelihood-ratio statistic against the baseline is
    $LR = 22.08$ on one degree of freedom; the $5\%$
    critical value is $3.84$.}
  \label{tab:jrssa_muC_2025_comparison}
  \begin{tabular}{lcc}
    \toprule
    & Baseline (20p) & With $\mu_C$ (21p) \\
    \midrule
    log-likelihood              & $-85.515$ & $-74.476$ \\
    \midrule
    $\beta_1$ (land fertilisation slope)
                                & $6.416$ $(0.286)$ & $6.909$ $(0.204)$ \\
    $\beta_2$ (ocean fertilisation slope)
                                & $7.174$ $(0.298)$ & $6.272$ $(0.189)$ \\
    $\phi_3$ (ocean AR(1))      & $0.9906$ $(0.0280)$ & $0.6063$ $(0.1192)$ \\
    $c_1$ (smoothed, last yr)   & $-6.6522$ $(0.0715)$ & $-7.2438$ $(0.0704)$ \\
    $c_2$ (smoothed, last yr)   & $-6.6629$ $(0.1022)$ & $-5.8956$ $(0.0391)$ \\
    $\mu_C$                     & n.a. & $-0.5874$ $(0.0884)$ \\
    \midrule
    mean$(S^{LND}_{obs} - S^{LND*})$ (GtC/yr)
                                & $-0.0202$ & $+0.0023$ \\
    mean$(S^{OCN}_{obs} - S^{OCN*})$ (GtC/yr)
                                & $-0.4361$ & $-0.0010$ \\
    \bottomrule
  \end{tabular}
\end{table}

The augmented fit also pulls the fertilisation slopes in the direction of the
$2017$--$2024$ range: $\hat\beta_1$ rises from $6.42$ to $6.91$
(the $2017$--$2024$ range is $[7.30, 9.10]$), and
$\hat\beta_2$ falls from $7.17$ to $6.27$ (the $2017$--$2024$ range
is $[5.24, 5.67]$).  The baseline $(\beta_1, \beta_2)$ shift in 2025
is therefore partly real and partly an artifact of the constrained
budget identity.  Table~\ref{tab:jrssa_muC_2025_comparison} compares the two fits with and without constant
on the 1959--2016 sample.  A constant is the appropriate augmentation on
this sample.  The reported imbalance has no significant trend on
$1959$--$2016$ ($P=0.19$), so a trend term in the budget equation is not
identified by the data on which the model is estimated.
Appendix~\ref{app:muC_full} reports the fit of the MDS-GCB model (with $\mu_C$) on the full $1959$--$2024$ sample.

The same augmentation applied to every vintage tests whether the constant
was needed before 2025.  Table~\ref{tab:jrssa_muC_vintages} reports
$\hat\mu_C$, its standard error, and the likelihood-ratio statistic
against the corresponding baseline fit, for all nine vintages on the
common sample.  For 2017 to 2024 the estimate lies between $-0.149$ and
$+0.043$~GtC/yr, and the restriction $\mu_C = 0$ is not rejected at the
five per cent level in any vintage; the smallest $P$-value is $0.084$,
on 2019.  For 2025, $\hat\mu_C = -0.587$ $(0.088)$ and $LR = 22.08$ on
one degree of freedom.  The 2025 constant is four times the largest of
the eight preceding estimates.

The constant is inert on the vintages that do not need it.  Adding
$\mu_C$ leaves the fertilisation slopes of the 2017--2024 vintages
within their baseline ranges, $\hat\beta_1$ between $7.30$ and $9.08$
and $\hat\beta_2$ between $5.29$ and $5.66$, and leaves the ocean
persistence $\hat\phi_3$ between $0.53$ and $0.66$.  On 2025 the same
parameter moves $\hat\beta_1$ from $6.42$ to $6.91$, $\hat\beta_2$ from
$7.17$ to $6.27$, and $\hat\phi_3$ from $0.99$ to $0.61$.

\begin{table}[htbp]
  \centering
  \footnotesize
  \caption{Test of $\mu_C = 0$ across GCB vintages.  The MDS-GCB model is
    estimated with and without a constant $\mu_C$ in the budget-identity
    state equation, on the common $1959$--$2016$ sample.  $LR$ is twice
    the log-likelihood difference against the $20$-parameter baseline,
    distributed $\chi^2_1$ under the restriction; the $5\%$ critical
    value is $3.84$.  The standard error for $\hat\mu_C$ is from the
    regularised inverse Hessian.  $\hat\beta_1$, $\hat\beta_2$ and
    $\hat\phi_3$ are from the augmented fit.}
  \label{tab:jrssa_muC_vintages}
  \begin{tabular}{lrrrrrr}
    \toprule
    Vintage & $\hat\mu_C$ & (SE) & $LR$ & $P$ & $\hat\beta_1$ & $\hat\beta_2$ \\
    \midrule
    2017 & $+0.008$ & $(0.099)$ & $0.01$ & $0.935$ & $8.64$ & $5.45$ \\
    2018 & $-0.125$ & $(0.087)$ & $1.98$ & $0.160$ & $9.07$ & $5.29$ \\
    2019 & $-0.149$ & $(0.084)$ & $2.98$ & $0.084$ & $8.82$ & $5.55$ \\
    2020 & $+0.018$ & $(0.081)$ & $0.05$ & $0.832$ & $9.08$ & $5.41$ \\
    2021 & $-0.048$ & $(0.084)$ & $0.32$ & $0.571$ & $7.30$ & $5.53$ \\
    2022 & $+0.043$ & $(0.088)$ & $0.23$ & $0.632$ & $8.25$ & $5.66$ \\
    2023 & $+0.020$ & $(0.082)$ & $0.06$ & $0.805$ & $8.34$ & $5.56$ \\
    2024 & $-0.044$ & $(0.081)$ & $0.23$ & $0.629$ & $8.74$ & $5.61$ \\
    \bfseries{}2025 & $\mathbf{-0.587}$ & $\mathbf{(0.088)}$ & $\mathbf{22.08}$ & $\mathbf{<0.001}$ & $\mathbf{6.91}$ & $\mathbf{6.27}$ \\
    \bottomrule
  \end{tabular}
\end{table}

The constant is only a partial fix, and the cumulative-CO$_2$ side shows
why.  A single constant absorbs the \emph{mean} of a
budget discrepancy but not a possible drift in the budget equation over time.  
The 2025 revision introduces a level shift, but it turns out that even prior to 2025,
there has been a slowly drifting discrepancy between the observed atmospheric-CO$_2$ accumulation and the
budget-implied stock. This is a finding that the model-free first stage
of the audit could not show. The model carries this drift through the one
residual with the memory to hold it: the cumulative-CO$_2$ measurement
residual $X^{(1)}$, whose persistence $\phi_1$ jumps from 0.81 in 2023 to 0.99 in 2024, then to the
random-walk boundary $1.00$ in 2025.

On the $2017$--$2023$ vintages this residual is stationary with strong serial correlation (see Appendix~\ref{app:full_parameters}). 
Figure~\ref{fig:muC_X1_2025} shows the graphs for the baseline MDS-GCB model without constant in the budget equation. On the 2017--2024 vintages $X_1$ wanders by about $5$~GtC over the sample, a small fraction of the $\sim\!700$~GtC stock but a persistent, low-frequency
drift that a constant cannot represent. The vintages 2020 and 2024 already exhibit highly persistent excursions from the mean. 2025 takes off to over 10 GtC/yr. Introducing the constant in the budget equation pulls the excursion back to about 5 GtC/yr (not shown), but the serial correlation parameter remains at the random-walk boundary.

\begin{figure}[htbp]
  \centering
  \includegraphics[width=0.78\linewidth]{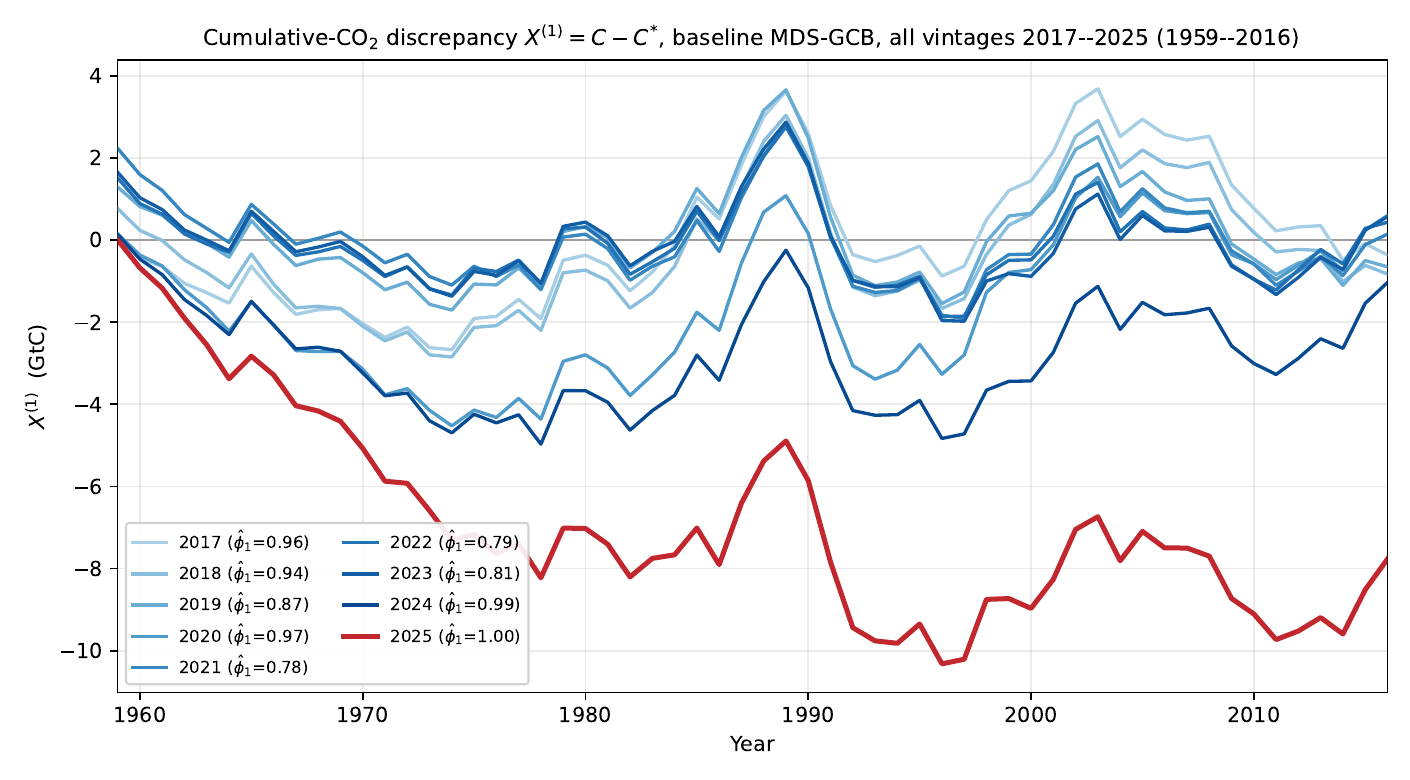}
  \caption{Smoothed cumulative-CO$_2$ measurement residual
    $X^{(1)} = C - C^{*}$ under the $\mu_C$-augmented MDS-GCB on the
    2025 vintage, common $1959$--$2016$ sample, with pointwise $95\%$
    smoothed-state band.}
  \label{fig:muC_X1_2025}
\end{figure}

\FloatBarrier

\section{Discussion}
\label{sec:discussion}

The two-part audit produces two complementary readings of the 2024 to
2025 vintage shift.  Part~1 (Section~\ref{sec:audit_panel}) reads
the shift directly off the reported data sets: the 2025 GCB
budget imbalance of $+0.606$~GtC/yr decomposes as an ensemble-mean
imbalance of $+0.411$, a land-sink adjustment gap of $+0.400$ (the RSS
bias correction introduced in the 2025 release), and a partially
offsetting ocean-sink adjustment gap of $-0.205$ (the underestimation
adjustment to the GOBM ensemble mean), each near zero in every earlier
vintage. An Oaxaca-Blinder decomposition of the ensemble-mean
shifts identifies specific drivers for LUC, land sink, and ocean sink ensembles.  

Part~2 (Section~\ref{sec:jrssa_vintages})
re-runs the audit through the MDS-GCB model of
\cite{BHK2023} with parameters estimated for every vintage 2017--2025. The results show that
the 2025 model-free shifts perturb the fertilisation slopes of the land and ocean sinks by roughly seven averaged standard errors each
against the 2017--2024 baseline, in opposite directions. The model-based analysis also shows
the slow build-up of a persistent drift in the budget equation across time,
first visible in 2024 and very clearly diagnosed in 2025.

The two readings are mutually consistent.  The 2025 land-sink
adjustment moves the published series away from the DGVM ensemble mean,
and the MDS-GCB structural model reads this as a reduced land
fertilisation slope; the ocean-side adjustment moves the published
series above the ocean ensemble mean, and the MDS-GCB model reads
this as an increased ocean fertilisation slope.

The 2025 release documents each methodological change in Table~3 of
\citet{GCB2025}, and gives each an evidence base that does not
refer to the budget imbalance.

For the land sink, Section~2.6.1 introduces the Replaced Sinks and
Sources (RSS) correction.  The DGVMs compute the sink on the
pre-industrial land surface, while a large part of that surface has
since been converted to cropland and pasture, which has a lower sink
capacity.  The resulting bias is established in \citet{GitzCiais2003},
\citet{Sitch2005}, \citet{Pongratz2009}, \citet{Gasser2020},
\citet{Obermeier2021} and \citet{Dorgeist2024}, and it is quantified for
the GCB ensemble at the level of plant functional types by
\citet{OSullivan2025}.  The correction reduces $S^{LND}$ by $19\%$
globally, $0.6$~GtC/yr over $2015$--$2024$.

For the ocean sink, Section~2.5 applies two corrections.  The fCO$_2$
products are recomputed at the temperature of the surface layer where
the gas exchange takes place rather than at the temperature at
measurement depth \citep{Dong2022}, which adds $0.18$~GtC/yr.  The GOBM
ensemble mean is scaled up by $10\%$, because of an assessment that the
models understate the sink by $10$ to $20$ per cent, based on
ocean-interior carbon estimates, atmospheric oxygen, and atmospheric
inversions, and supported by eddy-covariance measurements
\citep{Dong2024}.

For land-use emissions, Section~2.2 reports the transition of the
bookkeeping ensemble to BLUE, OSCAR and LUCE.  All three use transient
carbon densities, so that environmental change is allowed to affect
vegetation and soil carbon stocks.  \citet{HoughtonCastanho2023} does
not, and its data end in 2020, which would require extrapolation over
the last four years of the record.  It is dropped on those two grounds.

None of these derivations refers to the budget imbalance, and we take
each as given.  The audit measures their joint effect on the published
budget, which no single component derivation reports.

The 2025 GCB paper reports the budget imbalance and interprets it.
Section~3.9.1 states that the RSS correction reduces $S^{LND}$ by about
$40$~GtC over $1850$--$2024$, ``contributing to the increase of the
historical budget imbalance in comparison to GCB2024,'' and reports the
$1959$--$2024$ imbalance as $35$~GtC, seven per cent of total emissions.
Section~3.9.2 gives the same quantity as an annual mean of $0.5$~GtC/yr
and reads a small mean in the imbalance as ``evidence of a coherent
community understanding of the emissions and their partitioning on those
time scales.''  Our full-sample mean for the 2025 vintage,
$+0.54$~GtC/yr, agrees with theirs.

We take that criterion as given and supply the benchmark it requires.  A
mean of $0.5$~GtC/yr is small against the year-to-year variability of
the imbalance, which is of order $\pm 1$~GtC/yr.  It is not small
against the same statistic in the eight preceding vintages, which lies
within $\pm 0.17$~GtC/yr on a common sample.  On that comparison the
2025 vintage is the first for which the criterion of Section~3.9.2 is
not met.

Section~3.9.1 attributes the increase in the historical imbalance to the
RSS correction.  We put the RSS correction at $+0.400$~GtC/yr of the
$+0.606$~GtC/yr total on the common sample, with the removal of
H\&C2023 contributing a further $+0.195$ through the published
$E^{LUC}$ and the ocean-sink correction offsetting $-0.205$.  The model-based audit in
Section~\ref{sec:jrssa_vintages} adds a structural reading: the 2025
release with its substantial non-zero mean budget imbalance distorts
sink estimates and derived statistics unless a bias offset is added
to the concentrations equation, effectively re-introducing a
``missing sink'' that is estimated at $0.59$~GtC with standard
error $0.09$, if one imposes a constant. The model dynamics reveal, however,
that the tension in the budget equation has persistent dynamics.

A near-zero imbalance mean in the earlier vintages is consistent with
compensating biases in the components rather than with each component
being unbiased.  The test in Section~\ref{sec:audit_model_bim} does not
distinguish the two, and does not need to: whatever produced the
cancellation, it was sufficient for the budget identity to hold without
a constant in every vintage from 2017 to 2024.  It ceases to be
sufficient in 2025.

\subsection*{Implications for users of the GCB}

The audit produces four practical messages for researchers who use
the GCB on a recurring basis.

First, the carbon-budget identity no longer closes in mean on the
2025 vintage.  For every vintage from 2017 to 2024, the mean of the budget imbalance fell within an interval of $\pm 0.17$~GtC/yr on the
common $1959$--$2016$ sample, with confidence intervals covering
zero.  The 2025 vintage jumps to $+0.61$~GtC/yr, with a confidence
interval excluding zero under both i.i.d.\ and AR(1) assumptions.
Any quantity computed from the published series that relies on the
identity (the implied total sink, the sink rate computed from that
implied total sink rather than from the published sinks, attribution
that subtracts one component from another) now
depends on whether the user treats the imbalance as measurement
error to be ignored or as a real feature of the published budget.  The two treatments give different answers, and
users should report which convention they use.

Second, derived statistics that the wider literature reports,
notably the airborne fraction and the sink rate, are no longer
directly comparable across vintages without a fixed accounting
convention. Cross-vintage AF and SR time series stitched
together from multiple GCB releases inherit a discontinuity at
2025 whose size and sign depend on the user's choice of accounting.
Multi-vintage tracking of these statistics should be re-derived
under a single convention held constant.

Third, the 2025 release records each methodological change and reports
that the historical budget imbalance rises.  It does not carry that
through to the statistics derived from the budget.  The airborne
fraction and the sink rate both fall by about five per cent in 2025, and
neither effect is reported.  The three-step model-free audit
generalises to any future GCB vintage transition without
modification; the model-based audit similarly extends with the
$\mu_C$ augmentation on any vintage in which the published budget
identity does not close. Future adjustments may necessitate a time-varying $\mu_{C,t}$.

\subsection*{Limitations}

The audit is a within-dataset procedure: it compares the
published GCB numbers against either the underlying model ensembles (Part~1) or the MDS-GCB structural reading (Part~2), with
both reference points drawn from within the GCB itself.  It does
not engage independent benchmarks for the budget components, such
as atmospheric inversions for the sinks or alternative
land-cover proxies for LUC.  

The model-based part of the audit relies on the MDS-GCB specification of
\cite{BHK2023}, which imposes the carbon-budget identity
deterministically as a state equation on $C^{*}$.  The $\mu_C$
augmentation of Section~\ref{sec:audit_model_bim} does not repair the
cumulative-CO$_2$ residual.  Its persistence $\hat\phi_1$ rises from
$0.81$ on the 2023 vintage to $0.98$ on 2024 and reaches the
random-walk boundary on 2025, and it remains at that boundary in the
augmented fit.  A constant absorbs the mean of the discrepancy between
measured and budget-implied concentrations, not its drift.  We estimate
$\mu_C$ on all nine vintages and retain it where the data require it.  A
time-varying $\mu_{C,t}$ would represent that drift, which a constant
does not.

\section{Conclusion}
\label{sec:conclusion}

The Global Carbon Budget is the community reference for analyses of
the carbon cycle, and the reliability of those analyses depends on
the reliability of the GCB vintage being used.  Vintage-to-vintage
methodological updates can change the published numbers materially,
as the 2024 to 2025 transition demonstrated: the GCB budget imbalance
moved from approximately zero to $+0.61$~GtC/yr in a single release.

This paper conducted an audit in two stages. Part~1 read the evidence model-free, decomposing the 2025 budget imbalance into an ensemble-mean imbalance term and component-specific adjustment gaps between the ensemble means and the GCB values. The dominant signal is the `$+0.400$`~GtC/yr land-sink adjustment gap, attributable to the RSS bias correction, which the 2025 spreadsheet reports as a distinction between ``GCB'' and ``GCB unadjusted'' columns. A smaller ocean-sink adjustment gap of `$-0.205$`~GtC/yr runs in the opposite direction, the GCB ocean sink sitting above its ensemble mean and partly offsetting the land-sink gap; together with an ensemble-mean imbalance of `$+0.411$`~GtC/yr, these three terms reproduce the `$+0.606$`~GtC/yr GCB budget imbalance. An Oaxaca-Blinder decomposition of the ensemble-mean shifts identified three additional findings: the removal of the H\&C2023 bookkeeping model from the BK ensemble (a single-member composition effect on the LUC reference), a coherent ensemble-wide downward revision of the DGVM LUC ensemble on nearly all continuing members, and several newly added land-sink DGVMs with low sink values that pull the ensemble mean down.

Part~2 carried the audit through the four-series MDS-GCB
model of \cite{BHK2023}, augmented with the ENSO~3.4
\citep{HuangTBZSMLL2017}, NAO \citep{NOAA_NAO}, PDO \citep{NOAA_PDO},
and SAOD \citep{SatoHLR1993,KovilakamDTR2020} series in the sink
state equations, and estimated on every GCB vintage 2017--2025 on the
common 1959--2016 sample.  The land fertilisation slope $\beta_1$ in
2025 sits roughly seven averaged standard errors below the
2017--2024 cluster and the ocean fertilisation slope $\beta_2$ roughly seven
averaged standard errors above it in the opposite direction; the
MDS-GCB model reads the land-sink adjustment and
ocean-sink shift as a joint reweighting of the fertilisation slopes. The model also diagnoses
the slow build-up of a persistent drift in the budget equation across time,
first visible in 2024 and very crisp in 2025, where a non-stationary adjustment is necessary to capture the discrepancy
between measured changes in atmospheric concentrations and budget-implied concentrations.

Any quantities computed or derived from the GCB that rely on a closed budget equation, such as different expressions for the sink rate, need to take into account that the budget no longer closes on the full sample. The 2025 adjustment introduces a discontinuity in the sequence of vintages that is challenging to reconcile and may distort analyses that rely on historical data.

\clearpage
\addcontentsline{toc}{section}{Acknowledgements}
\section*{Acknowledgements}

We thank Pierre Friedlingstein, Glen Peters and Mike O'Sullivan for
their comments on an earlier version of this paper.
Section~\ref{sec:audit_step5} was added in response to a question raised
by Glen Peters.

\clearpage
\addcontentsline{toc}{section}{References}
\bibliographystyle{apalike}
\bibliography{references}

\clearpage
\appendix
\begin{center}
  {\Large\bfseries Appendix}
\end{center}
\vspace{1em}

\section{Oaxaca-Blinder decompositions for all vintages}
\label{sec:appB}

This appendix reports the Oaxaca-Blinder decomposition
\eqref{eq:oaxaca} applied separately to each of the seven consecutive
GCB vintage transitions 2017--2018 through 2023--2024.  The 2024--2025 transition is reported in the main text
(Table~\ref{tab:audit_step4}, Section~\ref{sec:audit_step4}).

For each ensemble and each transition the decomposition is computed on
the common 1959--2016 sample.  We match models across vintages under
known renames (for example CABLE / CABLE-POP, CLM4.5(BGC) / CLM5.0 /
CLM6.0, CLASS-CTEM / CLASSIC, JULES / JULES-ES, OCN / OCNv2, and the
LPX-Bern bookkeeping versus LPX-Bern DGVM entries); members with
genuinely different names across vintages are treated as composition
events.  Tables~\ref{tab:appB_bk}--\ref{tab:appB_ocean} report
$\Delta \bar X$, the revision component on continuing members
$\Delta_{\mathrm{rev}}$, the inflow component from new members
$\Delta_{\mathrm{in}}$, and the outflow component from removed members
$-\Delta_{\mathrm{out}}$, for the bookkeeping LUC ensemble, the DGVM LUC
ensemble, the DGVM land-sink ensemble, and the ocean ensemble respectively.

The ensemble-mean shifts and individual components are small across
earlier transitions: no $|\Delta \bar X|$ exceeds $0.310$~GtC/yr
(ocean ensemble, 2020--2021) and no individual component of the decomposition exceeds
$0.283$~GtC/yr (also ocean ensemble, 2021--2022, $\Delta_{\mathrm{out}}$).
The 2024--2025 BK ensemble-mean shift of $+0.261$~GtC/yr is comparable
in magnitude to several earlier transitions, but only 2024--2025
combines a large $\Delta \bar X$ with a single-event composition
explanation (the removal of H\&C2023).  The 2024--2025 DGVM LUC
revision component of $-0.380$~GtC/yr is the largest single 
component across all eight transitions and is the strongest evidence
that the 2024--2025 transition is qualitatively different from the
preceding eight years of vintage history.

\begin{table}[h]
  \centering
  \caption{Appendix B. Oaxaca-Blinder decomposition of the BK ensemble-mean change across consecutive non-headline GCB vintages. Common $1959$--$2016$ sample. Components: revision on continuing members, composition inflow from new members, composition outflow from removed members. Decomposition identity $\Delta \bar X = \Delta_{\mathrm{rev}} + \Delta_{\mathrm{in}} - \Delta_{\mathrm{out}}$.}
  \label{tab:appB_bk}
  \begin{tabular}{lrrrrrrr}
    \toprule
    Transition & $N_v\!\to\! N_w$ & $N_{\mathrm{in}}$ & $N_{\mathrm{out}}$ & $\Delta \bar X$ & $\Delta_{\mathrm{rev}}$ & $\Delta_{\mathrm{in}}$ & $-\Delta_{\mathrm{out}}$ \\
    \midrule
    2017-2018 & $2\!\to\!2$ & 0 & 0 & $+0.049$ & $+0.049$ & $+0.000$ & $-0.000$ \\
    2018-2019 & $2\!\to\!2$ & 0 & 0 & $+0.006$ & $+0.006$ & $+0.000$ & $-0.000$ \\
    2019-2020 & $2\!\to\!3$ & 1 & 0 & $+0.062$ & $+0.028$ & $+0.033$ & $-0.000$ \\
    2020-2021 & $3\!\to\!3$ & 0 & 0 & $-0.079$ & $-0.079$ & $+0.000$ & $-0.000$ \\
    2021-2022 & $3\!\to\!3$ & 0 & 0 & $+0.053$ & $+0.053$ & $+0.000$ & $-0.000$ \\
    2022-2023 & $3\!\to\!3$ & 1 & 1 & $+0.082$ & $+0.090$ & $-0.268$ & $+0.261$ \\
    2023-2024 & $3\!\to\!4$ & 1 & 0 & $+0.036$ & $-0.035$ & $+0.072$ & $-0.000$ \\
    \bottomrule
  \end{tabular}
\end{table}

\begin{table}[h]
  \centering
  \caption{Appendix B. Oaxaca-Blinder decomposition of the DGVM LUC ensemble-mean change across consecutive non-headline GCB vintages. Common $1959$--$2016$ sample. Components: revision on continuing members, composition inflow from new members, composition outflow from removed members. Decomposition identity $\Delta \bar X = \Delta_{\mathrm{rev}} + \Delta_{\mathrm{in}} - \Delta_{\mathrm{out}}$.}
  \label{tab:appB_dgvm_luc}
  \begin{tabular}{lrrrrrrr}
    \toprule
    Transition & $N_v\!\to\! N_w$ & $N_{\mathrm{in}}$ & $N_{\mathrm{out}}$ & $\Delta \bar X$ & $\Delta_{\mathrm{rev}}$ & $\Delta_{\mathrm{in}}$ & $-\Delta_{\mathrm{out}}$ \\
    \midrule
    2017-2018 & $12\!\to\!16$ & 6 & 2 & $+0.229$ & $+0.275$ & $-0.063$ & $+0.017$ \\
    2018-2019 & $16\!\to\!16$ & 1 & 1 & $-0.059$ & $-0.054$ & $-0.072$ & $+0.073$ \\
    2019-2020 & $16\!\to\!17$ & 3 & 2 & $+0.124$ & $+0.089$ & $+0.021$ & $+0.014$ \\
    2020-2021 & $17\!\to\!17$ & 0 & 0 & $-0.096$ & $-0.096$ & $+0.000$ & $-0.000$ \\
    2021-2022 & $17\!\to\!16$ & 0 & 1 & $+0.022$ & $-0.011$ & $+0.000$ & $+0.034$ \\
    2022-2023 & $16\!\to\!20$ & 5 & 1 & $+0.153$ & $+0.087$ & $+0.022$ & $+0.045$ \\
    2023-2024 & $20\!\to\!20$ & 1 & 1 & $-0.029$ & $-0.026$ & $-0.010$ & $+0.006$ \\
    \bottomrule
  \end{tabular}
\end{table}

\begin{table}[h]
  \centering
  \caption{Appendix B. Oaxaca-Blinder decomposition of the DGVM $S_{\mathrm{LND}}$ ensemble-mean change across consecutive non-headline GCB vintages. Common $1959$--$2016$ sample. Components: revision on continuing members, composition inflow from new members, composition outflow from removed members. Decomposition identity $\Delta \bar X = \Delta_{\mathrm{rev}} + \Delta_{\mathrm{in}} - \Delta_{\mathrm{out}}$.}
  \label{tab:appB_dgvm_lnd}
  \begin{tabular}{lrrrrrrr}
    \toprule
    Transition & $N_v\!\to\! N_w$ & $N_{\mathrm{in}}$ & $N_{\mathrm{out}}$ & $\Delta \bar X$ & $\Delta_{\mathrm{rev}}$ & $\Delta_{\mathrm{in}}$ & $-\Delta_{\mathrm{out}}$ \\
    \midrule
    2017-2018 & $15\!\to\!16$ & 3 & 2 & $-0.162$ & $-0.039$ & $-0.074$ & $-0.049$ \\
    2018-2019 & $16\!\to\!16$ & 1 & 1 & $-0.000$ & $+0.003$ & $-0.000$ & $-0.003$ \\
    2019-2020 & $16\!\to\!17$ & 3 & 2 & $+0.159$ & $+0.088$ & $+0.044$ & $+0.028$ \\
    2020-2021 & $17\!\to\!17$ & 0 & 0 & $-0.201$ & $-0.201$ & $+0.000$ & $-0.000$ \\
    2021-2022 & $17\!\to\!16$ & 0 & 1 & $+0.102$ & $+0.078$ & $+0.000$ & $+0.024$ \\
    2022-2023 & $16\!\to\!20$ & 5 & 1 & $+0.040$ & $-0.014$ & $+0.020$ & $+0.034$ \\
    2023-2024 & $20\!\to\!20$ & 2 & 2 & $-0.054$ & $-0.057$ & $+0.017$ & $-0.014$ \\
    \bottomrule
  \end{tabular}
\end{table}

\begin{table}[h]
  \centering
  \caption{Appendix B. Oaxaca-Blinder decomposition of the Ocean ensemble-mean change across consecutive non-headline GCB vintages. Common $1959$--$2016$ sample. Components: revision on continuing members, composition inflow from new members, composition outflow from removed members. Decomposition identity $\Delta \bar X = \Delta_{\mathrm{rev}} + \Delta_{\mathrm{in}} - \Delta_{\mathrm{out}}$.}
  \label{tab:appB_ocean}
  \begin{tabular}{lrrrrrrr}
    \toprule
    Transition & $N_v\!\to\! N_w$ & $N_{\mathrm{in}}$ & $N_{\mathrm{out}}$ & $\Delta \bar X$ & $\Delta_{\mathrm{rev}}$ & $\Delta_{\mathrm{in}}$ & $-\Delta_{\mathrm{out}}$ \\
    \midrule
    2017-2018 & $10\!\to\!9$ & 1 & 2 & $+0.029$ & $+0.021$ & $-0.004$ & $+0.012$ \\
    2018-2019 & $9\!\to\!12$ & 4 & 1 & $+0.008$ & $-0.026$ & $+0.003$ & $+0.031$ \\
    2019-2020 & $12\!\to\!14$ & 4 & 2 & $+0.033$ & $-0.017$ & $+0.046$ & $+0.004$ \\
    2020-2021 & $14\!\to\!16$ & 5 & 3 & $+0.310$ & $+0.169$ & $+0.178$ & $-0.036$ \\
    2021-2022 & $16\!\to\!18$ & 10 & 8 & $-0.009$ & $+0.076$ & $+0.198$ & $-0.283$ \\
    2022-2023 & $18\!\to\!18$ & 6 & 6 & $-0.021$ & $-0.012$ & $+0.128$ & $-0.137$ \\
    2023-2024 & $18\!\to\!19$ & 4 & 3 & $+0.071$ & $+0.055$ & $+0.146$ & $-0.130$ \\
    \bottomrule
  \end{tabular}
\end{table}

\section{Model specification and data}
\label{sec:jrssa_brief}

\subsection{Specification}
\label{sec:jrssa_spec}

The four measurement series are the cumulative atmospheric CO$_2$ stock
$C_t$ (constructed by accumulating $G^{\mathrm{ATM}}_t$ from
a fixed 1959 baseline), anthropogenic emissions $E_t = E^{FF}_t + E^{LUC}_t$ (vintages 2017--2019) 
and $E_t = E^{FF}_t + E^{LUC}_t - S_t^{CARB}$ (vintages 2020-2025), the
GCB land sink $S^{LND}_t$, and the GCB ocean sink
$S^{OCN}_t$.  These four series are measured by the latent budget plus
mean-zero residual processes,
\begin{align}
  C_t                &= C^{*}_t + X^{(1)}_t, \\
  S^{LND}_t          &= S^{LND*}_t + \eta^{(2)}_t, \\
  S^{OCN}_t          &= S^{OCN*}_t + X^{(3)}_t, \\
  E_t                &= E^{*}_t + \varepsilon^{(E)}_t,
  \label{eq:jrssa_measurement}
\end{align}
where $X^{(1)}_t$ and $X^{(3)}_t$ are AR(1) processes driven by
zero-mean Gaussian innovations $\eta^{(1)}_t$ and $\eta^{(3)}_t$,
$\eta^{(2)}_t$ is a zero-mean Gaussian innovation entering the
land-sink measurement equation directly, and $\varepsilon^{(E)}_t$ is
a zero-mean Gaussian measurement noise on total emissions with
variance $\sigma^2_{\varepsilon_E}$.  The innovation pairs
$(\eta^{(1)}_t, \eta^{(2)}_t)$ and $(\eta^{(1)}_t, \eta^{(3)}_t)$ have
free correlations $r_{12}$ and $r_{13}$ respectively; the remaining
correlations are restricted to zero.

The latent state vector contains the atmospheric CO$_2$ stock $C^{*}_t$,
its annual increment $G^{ATM*}_t$, the land sink $S^{LND*}_t$, the
ocean sink $S^{OCN*}_t$, anthropogenic emissions $E^{*}_t$, the two
AR(1) measurement residuals $X^{(1)}_t$ and $X^{(3)}_t$, an AR(1)
residual $X^{(E)}_t$ on emissions, and three constant states $c_1, c_2, d$.  The budget identity enters as an exact
state equation,
\begin{equation}
  C^{*}_{t+1} - C^{*}_t
    = X^{(E)}_t + E^{*}_t + d - S^{LND*}_{t+1} - S^{OCN*}_{t+1}.
  \label{eq:jrssa_budget}
\end{equation}
Combined with the sink equations \eqref{eq:jrssa_sink_lnd}--\eqref{eq:jrssa_sink_ocn}
below and the random-walk-with-drift transition~\eqref{eq:jrssa_emissions}
for $E^{*}$, this closes the transition for the four-state vector
$(C^{*}, S^{LND*}, S^{OCN*}, E^{*})$. Substituting the sink equations
into~\eqref{eq:jrssa_budget} yields a single linear equation in
$C^{*}_{t+1}$, which we solve and back-substitute to recover
$S^{LND*}_{t+1}$ and $S^{OCN*}_{t+1}$.  The two sinks are linear in
$C^{*}_t$ with model-wide climate and volcanic loadings,
\begin{align}
  S^{LND*}_{t+1}
    &= c_1 + \tfrac{\beta_1}{C_{1750}}\, C^{*}_{t+1}
    + b_1^{\mathrm{ENSO}}\,\mathrm{ENSO}_{t+1}
    + b_1^{\mathrm{NAO}}\,\mathrm{NAO}_{t+1}
    + b_1^{\mathrm{PDO}}\,\mathrm{PDO}_{t+1}
    + b_1^{\mathrm{SAOD}}\,\mathrm{SAOD}_{t+1},
    \label{eq:jrssa_sink_lnd}\\
  S^{OCN*}_{t+1}
    &= c_2 + \tfrac{\beta_2}{C_{1750}}\, C^{*}_{t+1}
    + b_2^{\mathrm{ENSO}}\,\mathrm{ENSO}_{t+1}
    + b_2^{\mathrm{NAO}}\,\mathrm{NAO}_{t+1}
    + b_2^{\mathrm{PDO}}\,\mathrm{PDO}_{t+1}
    + b_2^{\mathrm{SAOD}}\,\mathrm{SAOD}_{t+1},
    \label{eq:jrssa_sink_ocn}
\end{align}
with $C_{1750} = 593.43$~GtC the pre-industrial reference stock.  Emissions follow a random walk with drift, with an AR(1)
residual $X^{(E)}_t$,
\begin{equation}
  E^{*}_{t+1} = E^{*}_t + X^{(E)}_t + d, \qquad
  X^{(E)}_{t+1} = \phi_E X^{(E)}_t + \eta^{(E)}_{t+1}.
  \label{eq:jrssa_emissions}
\end{equation}
The three constants $c_1, c_2, d$ are estimated as constant latent
states, following \cite{DurbinKoopman2012}, Ch.~4.  The coefficients
$\beta_1$ and $\beta_2$ are the land and ocean fertilisation slopes:
each loads on the ratio of the current atmospheric CO$_2$ stock to its
pre-industrial level $C_{1750}$, so that the sink strengthens as
concentrations rise.  The four climate covariates carry, respectively,
the El Ni\~no--Southern Oscillation ($\mathrm{ENSO}\,3.4$), the North
Atlantic Oscillation ($\mathrm{NAO}$), the Pacific Decadal Oscillation
($\mathrm{PDO}$), and volcanic forcing through stratospheric aerosol
optical depth ($\mathrm{SAOD}$); Section~\ref{sec:jrssa_climate_data}
describes these data and their sources.  The state vector has dimension
eleven, with elements
($C^{*}, G^{ATM*}, S^{LND*}, S^{OCN*}, E^{*}, X^{(1)}, X^{(3)},
X^{(E)}, c_1, c_2, d$), and the structural parameter vector has
$20$ elements: the two fertilisation slopes $\beta_1, \beta_2$ just
described; the AR(1) persistences $\phi_1, \phi_3$ for $X^{(1)},
X^{(3)}$ and $\phi_E$ for $X^{(E)}$; four state-disturbance variances
$\sigma^2_{\eta_1}, \sigma^2_{\eta_2}, \sigma^2_{\eta_3},
\sigma^2_{\eta_E}$; the measurement-noise variance
$\sigma^2_{\varepsilon_E}$; the two innovation correlations $r_{12},
r_{13}$; and the eight climate loadings.

\subsection{Climate-driver data}
\label{sec:jrssa_climate_data}

The four climate-driver series that enter the sink state
equations~\eqref{eq:jrssa_sink_lnd}--\eqref{eq:jrssa_sink_ocn} are
aggregated to annual frequency in two different ways depending on the
phase relationship between the index and the annual carbon flux.
ENSO and NAO use the literature-standard windows centred on the
boreal winter peak that leads the next year's carbon response; PDO
and SAOD, both slower-varying, use calendar-year averaging.

\paragraph{ENSO~3.4.}  The Niño~3.4 sea-surface-temperature anomaly is
the NOAA Climate Prediction Center (CPC) monthly index constructed from the
Extended Reconstructed Sea Surface Temperature dataset version~5
\citep[ERSSTv5;][]{HuangTBZSMLL2017}.  The CPC distributes the
area-averaged Niño~3.4 SST and anomalies in a single ASCII table; the
anomaly column ANOM is the climatological anomaly relative to the
1991--2020 base period.  We aggregate the monthly anomalies to annual
frequency using a 12-month window from September of year $t-1$ through
August of year $t$, a four-month lag relative to the calendar year
that aligns the index with the lagged response of atmospheric CO$_2$
growth to ENSO.  This window matches the standard finding
that ENSO leads the atmospheric CO$_2$ growth rate by several months,
documented across the carbon-cycle literature
\citep{Zeng2005,WangTAR2013,LiuBR2017,WangCFGB2023}.  This series replaces the Southern Oscillation Index used in
the original specification of \cite{BHK2023}.

\paragraph{NAO.}  The North Atlantic Oscillation index is the
station-based monthly series distributed by the NOAA CPC \citep{NOAA_NAO}, constructed from sea-level-pressure
differences between the Azores and Iceland.  The annual value is the
winter (DJFM) mean of \cite{HurrellNAO1995}: the average of December
of year $t-1$ and January, February, and March of year $t$.  Winter
DJFM is the climatologically meaningful aggregation of the NAO and
the standard convention in the carbon-sink literature
\citep[e.g.,][]{BastosEA2016}.

\paragraph{PDO.}  The Pacific Decadal Oscillation index, in the form
introduced by \cite{MantuaHZWF1997}, is the ERSSTv5-based monthly
series from NOAA's National Centers for Environmental Information
\citep{NOAA_PDO}.  Like ENSO~3.4 it is constructed from ERSSTv5 sea
surface temperatures, so the two indices share a common climatology.
The PDO varies on decadal time scales by construction, so the choice
of 12-month aggregation window is immaterial; we use the calendar-year
mean.

\paragraph{SAOD.}  The stratospheric aerosol optical depth series
combines two sources.  The pre-satellite period 1959--1978 is from
NASA GISS \citep{SatoHLR1993}, which provides monthly SAOD at 550~nm
on a 24-band latitude grid; we compute the area-weighted global mean
using $\cos(\mathrm{lat})$ weights, then take a calendar-year mean.
The satellite era 1979 onwards is from the Global Space-based
Stratospheric Aerosol Climatology
\citep[GloSSAC v2.22;][]{KovilakamDTR2020}, which reports monthly SAOD
at four wavelengths on a finer 32-band latitude grid; we use the
525~nm band, apply the same $\cos(\mathrm{lat})$ weighting, and take a
calendar-year mean.  The two sources are spliced at 1979 with no
wavelength correction; the constant scale factor between the GISS
550~nm and GloSSAC 525~nm bands is absorbed into the
$b^{\mathrm{SAOD}}$ coefficients without affecting the AF or sink-rate
point estimates.  The Pinatubo eruption of June 1991 produces the
largest annual peak in the spliced series.  The volcanic forcing
itself decays over roughly $1.5$--$2$ years in the
stratosphere, so the calendar-year average captures the full
post-eruption signature; \cite{MercadoBHJC2009} document the
corresponding $1$--$2$-year enhancement of the global land carbon
sink.

The replication package contains full documentation.

\section{Full MDS-GCB parameter listing}
\label{app:full_parameters}

The MDS-GCB model of Section~\ref{sec:jrssa_brief} has $20$ free
parameters in total: two fertilisation slopes ($\beta_1, \beta_2$), three
persistence parameters ($\phi_1, \phi_3, \phi_E$), four
state-disturbance variances ($\sigma^2_{\eta_1}, \sigma^2_{\eta_2},
\sigma^2_{\eta_3}, \sigma^2_{\eta_E}$), one measurement-noise
variance ($\sigma^2_{\varepsilon_E}$), two innovation correlations
($r_{12}, r_{13}$), and eight climate state-equation loadings ($b_1,
b_2$ on each of ENSO, NAO, PDO, SAOD).  Three constants
$c_1, c_2, d$ are estimated as constant latent states and are reported
as the smoothed state values at the final
observation, with standard errors from the diagonal of the
smoothed-state covariance at that time.

Table~\ref{tab:jrssa_full_vintages} collects all 20 free parameters
plus the three constants for each GCB vintage 2017--2025
on the common 1959--2016 sample.  Standard errors in parentheses
come from the delta method applied to the regularised inverse of the
central-difference Hessian at the MLE.
The headline level-related block ($\beta_1, \beta_2, c_1, c_2, d$) is
also reported in Table~\ref{tab:jrssa_muC_2025_comparison} of
Section~\ref{sec:jrssa_vintages_struct}; the appendix repeats those
rows for completeness alongside the variance, persistence, and
climate-loading rows.

\begin{landscape}
\begin{table}[p]
  \centering
  \scriptsize
  \caption{Full MDS-GCB parameter table across GCB vintages 2017--2025, on the common 1959--2016 sample. Display units with delta-method standard errors in parentheses. The constants $c_1, c_2, d$ in the last three rows are smoothed at the final observation with standard errors from the smoothed-state covariance.}
  \label{tab:jrssa_full_vintages}
  \begin{tabular}{lccccccccc}
    \toprule
    Parameter & 2017 & 2018 & 2019 & 2020 & 2021 & 2022 & 2023 & 2024 & 2025 \\
    \midrule
    $\beta_1$ & $+8.649$ (0.304) & $+8.845$ (0.247) & $+8.705$ (0.427) & $+9.104$ (0.296) & $+7.296$ (0.233) & $+8.256$ (0.270) & $+8.346$ (0.228) & $+8.702$ (0.259) & $+6.416$ (0.286) \\
    $\beta_2$ & $+5.455$ (0.245) & $+5.240$ (0.236) & $+5.529$ (0.296) & $+5.420$ (0.202) & $+5.525$ (0.218) & $+5.666$ (0.213) & $+5.563$ (0.190) & $+5.577$ (0.217) & $+7.174$ (0.298) \\
    $\phi_1$ & $+0.957$ (0.020) & $+0.944$ (0.029) & $+0.871$ (0.048) & $+0.974$ (0.017) & $+0.780$ (0.030) & $+0.788$ (0.036) & $+0.807$ (0.050) & $+0.989$ (0.005) & $+1.000$ (0.000) \\
    $\phi_3$ & $+0.658$ (0.114) & $+0.583$ (0.122) & $+0.663$ (0.111) & $+0.527$ (0.130) & $+0.635$ (0.117) & $+0.573$ (0.126) & $+0.529$ (0.127) & $+0.528$ (0.125) & $+0.991$ (0.028) \\
    $\phi_E$ & $+0.511$ (0.212) & $+0.501$ (0.228) & $+0.466$ (0.230) & $+0.446$ (0.229) & $+0.128$ (0.180) & $+0.331$ (0.287) & $+0.308$ (0.251) & $+0.274$ (0.226) & $+0.188$ (0.150) \\
    $\sigma^2_{\eta_1}$ & $+0.460$ (0.060) & $+0.458$ (0.075) & $+0.442$ (0.071) & $+0.504$ (0.074) & $+0.390$ (0.056) & $+0.369$ (0.051) & $+0.378$ (0.066) & $+0.471$ (0.058) & $+0.562$ (0.113) \\
    $\sigma^2_{\eta_2}$ & $+0.457$ (0.081) & $+0.358$ (0.064) & $+0.329$ (0.067) & $+0.237$ (0.044) & $+0.340$ (0.065) & $+0.390$ (0.078) & $+0.330$ (0.057) & $+0.315$ (0.060) & $+0.171$ (0.034) \\
    $\sigma^2_{\eta_3}$ & $+0.007$ (0.001) & $+0.008$ (0.002) & $+0.006$ (0.001) & $+0.006$ (0.001) & $+0.006$ (0.001) & $+0.007$ (0.001) & $+0.009$ (0.002) & $+0.008$ (0.001) & $+0.008$ (0.002) \\
    $\sigma^2_{\eta_E}$ & $+0.012$ (0.006) & $+0.013$ (0.007) & $+0.011$ (0.006) & $+0.011$ (0.006) & $+0.026$ (0.007) & $+0.013$ (0.008) & $+0.016$ (0.009) & $+0.022$ (0.009) & $+0.032$ (0.007) \\
    $r_{12}$ & $-0.284$ (0.098) & $-0.285$ (0.089) & $-0.287$ (0.133) & $-0.241$ (0.112) & $-0.299$ (0.091) & $-0.371$ (0.109) & $-0.355$ (0.103) & $-0.359$ (0.091) & $-0.278$ (0.290) \\
    $r_{13}$ & $+0.196$ (0.142) & $+0.128$ (0.148) & $+0.147$ (0.144) & $+0.117$ (0.153) & $-0.000$ (0.143) & $-0.008$ (0.140) & $+0.019$ (0.142) & $-0.046$ (0.144) & $+0.093$ (0.200) \\
    $\sigma^2_{\varepsilon_E}$ & $+0.007$ (0.002) & $+0.007$ (0.003) & $+0.007$ (0.003) & $+0.006$ (0.002) & $+0.001$ (0.002) & $+0.006$ (0.003) & $+0.005$ (0.003) & $+0.006$ (0.003) & $+0.001$ (0.001) \\
    $b_1^{\mathrm{ENSO}}$ & $-1.040$ (0.130) & $-1.084$ (0.121) & $-1.020$ (0.119) & $-0.882$ (0.105) & $-0.975$ (0.110) & $-1.008$ (0.119) & $-0.982$ (0.114) & $-0.987$ (0.118) & $-0.757$ (0.106) \\
    $b_2^{\mathrm{ENSO}}$ & $+0.033$ (0.016) & $+0.050$ (0.018) & $+0.041$ (0.015) & $+0.051$ (0.017) & $+0.054$ (0.016) & $+0.049$ (0.018) & $+0.056$ (0.020) & $+0.055$ (0.019) & $+0.033$ (0.016) \\
    $b_1^{\mathrm{NAO}}$ & $+0.150$ (0.209) & $+0.160$ (0.194) & $+0.135$ (0.186) & $+0.172$ (0.168) & $+0.138$ (0.177) & $+0.226$ (0.188) & $+0.181$ (0.183) & $+0.176$ (0.190) & $+0.128$ (0.152) \\
    $b_2^{\mathrm{NAO}}$ & $+0.053$ (0.028) & $+0.061$ (0.031) & $+0.037$ (0.026) & $+0.046$ (0.029) & $+0.042$ (0.027) & $+0.042$ (0.030) & $+0.053$ (0.033) & $+0.034$ (0.031) & $+0.010$ (0.026) \\
    $b_1^{\mathrm{PDO}}$ & $+0.032$ (0.109) & $+0.050$ (0.102) & $+0.046$ (0.101) & $+0.071$ (0.087) & $+0.009$ (0.090) & $+0.000$ (0.094) & $+0.036$ (0.095) & $+0.069$ (0.098) & $+0.069$ (0.080) \\
    $b_2^{\mathrm{PDO}}$ & $+0.022$ (0.017) & $+0.016$ (0.018) & $+0.032$ (0.016) & $+0.039$ (0.017) & $+0.038$ (0.016) & $+0.048$ (0.018) & $+0.054$ (0.019) & $+0.050$ (0.019) & $+0.035$ (0.017) \\
    $b_1^{\mathrm{SAOD}}$ & $+13.558$ (4.188) & $+12.333$ (3.869) & $+12.167$ (3.821) & $+8.794$ (3.439) & $+12.890$ (3.621) & $+14.571$ (3.809) & $+13.307$ (3.690) & $+11.855$ (3.767) & $+5.738$ (3.998) \\
    $b_2^{\mathrm{SAOD}}$ & $+2.470$ (0.808) & $+2.769$ (0.887) & $+2.113$ (0.752) & $+2.428$ (0.820) & $+1.834$ (0.773) & $+2.134$ (0.866) & $+2.426$ (0.911) & $+1.914$ (0.884) & $+2.362$ (0.816) \\
    \midrule
    $c_1$ & $-9.021$ (0.064) & $-9.324$ (0.055) & $-9.113$ (0.041) & $-9.522$ (0.066) & $-7.459$ (0.032) & $-8.684$ (0.033) & $-8.705$ (0.033) & $-9.177$ (0.081) & $-6.652$ (0.071) \\
    $c_2$ & $-5.255$ (0.030) & $-4.926$ (0.027) & $-5.284$ (0.027) & $-5.186$ (0.026) & $-5.211$ (0.026) & $-5.343$ (0.025) & $-5.193$ (0.024) & $-5.177$ (0.033) & $-6.663$ (0.102) \\
    $d$   & $+0.126$ (0.030) & $+0.123$ (0.031) & $+0.128$ (0.026) & $+0.123$ (0.026) & $+0.108$ (0.025) & $+0.117$ (0.023) & $+0.115$ (0.024) & $+0.114$ (0.027) & $+0.112$ (0.030) \\
    \bottomrule
  \end{tabular}
\end{table}
\end{landscape}

The full code base (model class and data loaders, the
BFGS/Nelder-Mead numerical optimisation, and the regularised Hessian) is provided as a
Python replication package alongside this paper.

\section{The $\mu_C$ model on the full $1959$--$2024$ sample}
\label{app:muC_full}

The $\mu_C$-augmented MDS-GCB of Section~\ref{sec:audit_model_bim} is
estimated in the main text on the common $1959$--$2016$ sample, for
comparability with the $2017$--$2024$ vintages.  This appendix reports
the same model estimated on the 2025 vintage over its full available
sample $1959$--$2024$ ($T = 66$).  The purpose is completeness and a demonstration how the
drifting budget discrepancy discussed in
Section~\ref{sec:audit_model_bim} develops once the recent,
high-imbalance years are included.

On the full sample the fertilisation slopes are
$\hat\beta_1 = 5.77$ and $\hat\beta_2 = 5.69$, the ocean-sink
persistence is $\hat\phi_3 = 0.65$ (well inside the stationary range),
and the budget constant is $\hat\mu_C = -0.51$~GtC/yr.  The
cumulative-CO$_2$ persistence $\phi_1$ again sits at the random-walk
boundary $+1.00$.  Adding the $2017$--$2024$ years pulls $\beta_2$ down
to the top of the $2017$--$2024$ cluster range and leaves the constant
close to its $1959$--$2016$ value, so the reading of the ocean sink is stable
across the two windows.

\begin{figure}[htbp]
  \centering
  \includegraphics[width=\linewidth]{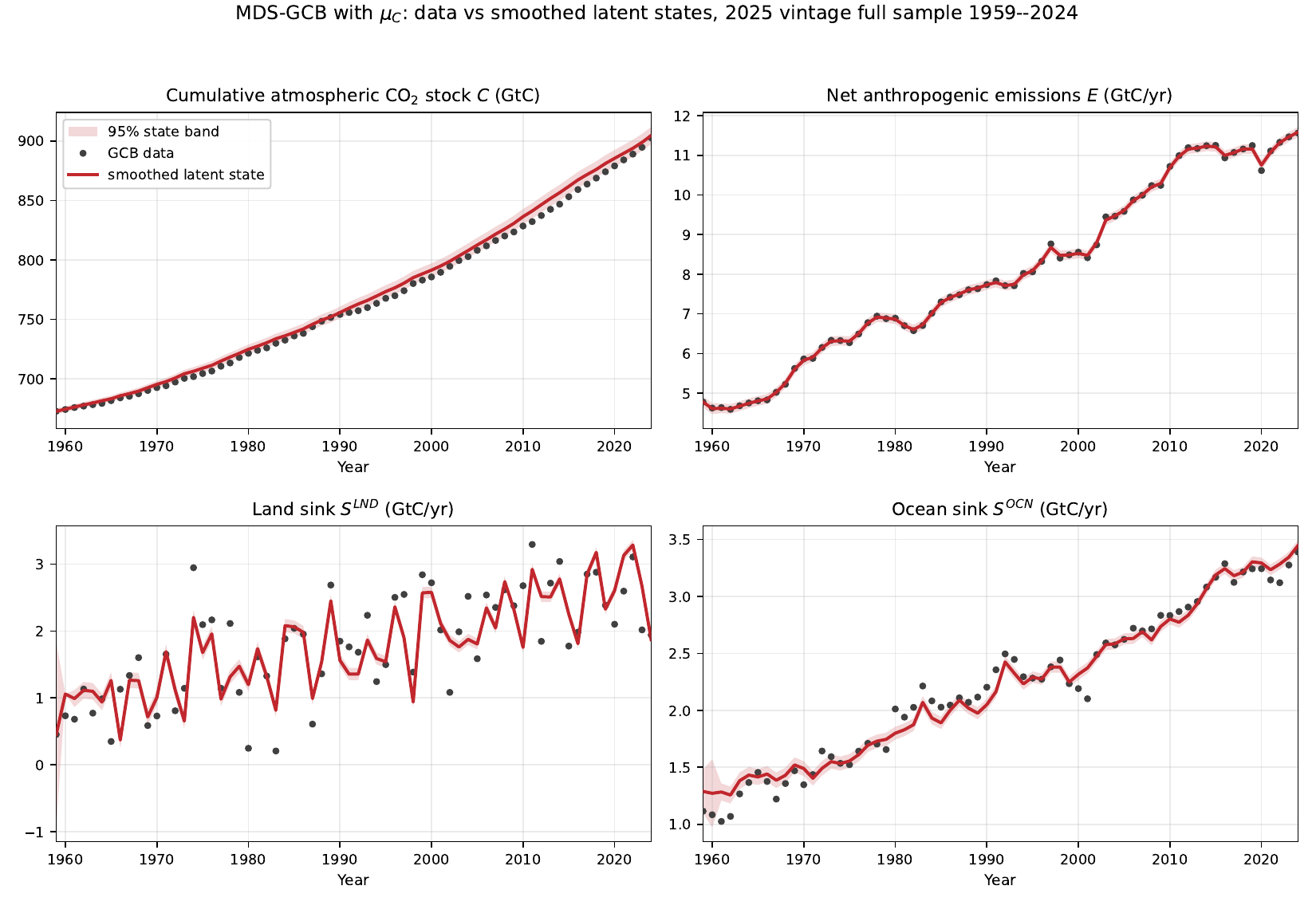}
  \caption{Data (grey) versus smoothed latent states (red) with
    pointwise $95\%$ smoothed-state bands, for the $\mu_C$-augmented
    MDS-GCB estimated on the 2025 vintage over the full $1959$--$2024$
    sample.  Panels: cumulative atmospheric CO$_2$ stock $C$, net
    anthropogenic emissions $E$, land sink $S^{LND}$, and ocean sink
    $S^{OCN}$.}
  \label{fig:muC_full_2x2}
\end{figure}

\begin{figure}[htbp]
  \centering
  \includegraphics[width=0.78\linewidth]{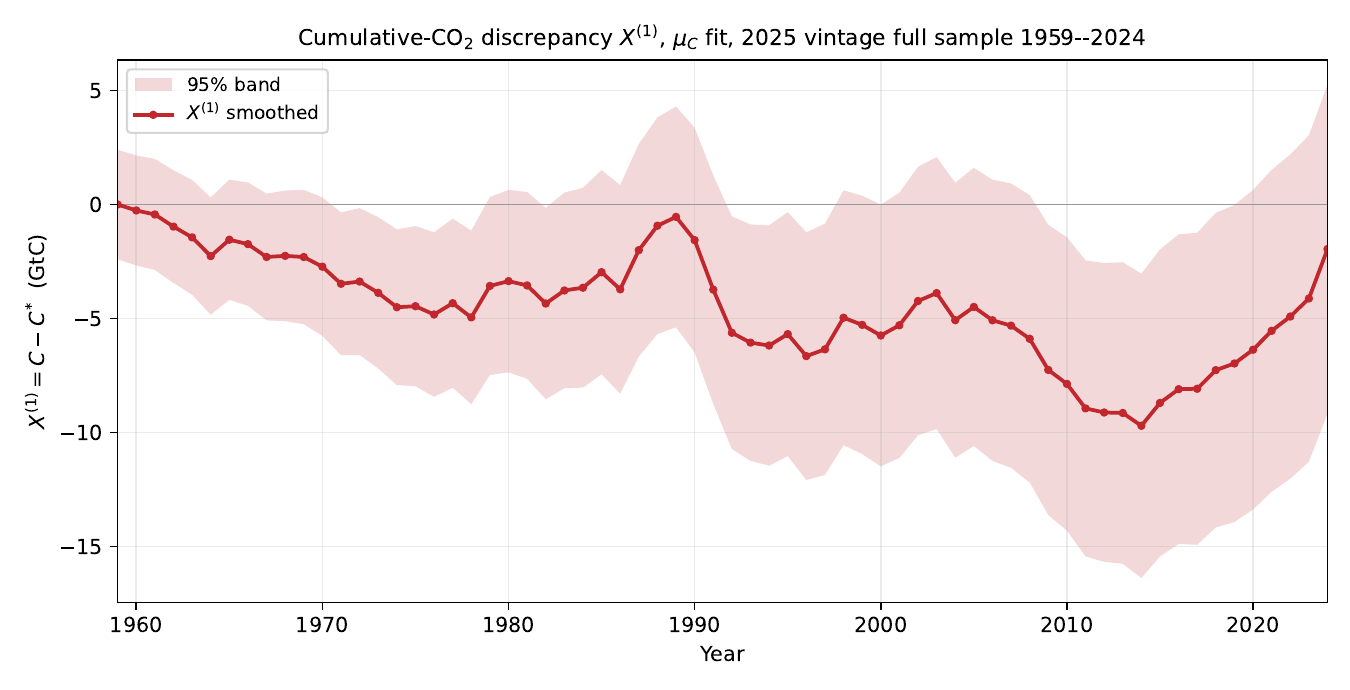}
  \caption{Smoothed cumulative-CO$_2$ measurement residual
    $X^{(1)} = C - C^{*}$ for the $\mu_C$-augmented MDS-GCB on the 2025
    vintage, full $1959$--$2024$ sample, with pointwise $95\%$
    smoothed-state band.  The drift reaches nearly $-10$~GtC, about
    twice the roughly $-4$~GtC excursion of the same $\mu_C$ fit on the
    $1959$--$2016$ window (Section~\ref{sec:audit_model_bim}).}
  \label{fig:muC_full_X1}
\end{figure}

Figure~\ref{fig:muC_full_2x2} plots the four measurement series against
their smoothed latent states.  With $\mu_C$ present, the smoothed land
and ocean sinks track their published series across the whole sample.
Figure~\ref{fig:muC_full_X1} plots the cumulative-CO$_2$ residual
$X^{(1)}$ for this fit.  Over the full sample it drifts to nearly
$-10$~GtC by the mid-2010s before recovering, about twice the
$\mu_C$-fit excursion on the $1959$--$2016$ window
(Section~\ref{sec:audit_model_bim}).  The larger
drift is consistent with the reading in
Section~\ref{sec:audit_model_bim}: the 2025 vintage introduces a
slowly drifting budget discrepancy, and the constant $\mu_C$ removes
only its mean, leaving the drift to accumulate in $X^{(1)}$.

\end{document}